\documentclass[aps,pra,amsmath,amssymb,reprint,groupedaddress]{revtex4-2}
\usepackage[utf8]{inputenc}
\usepackage[english]{babel} 
\usepackage[T1]{fontenc}
\usepackage{graphicx}
\usepackage{placeins}
\usepackage{amsmath,amssymb}
\usepackage[colorlinks=true,citecolor=red]{hyperref}		
\usepackage{natbib}
\usepackage{xcolor}
\usepackage{siunitx}
\usepackage{braket}

\newcommand\G{\mathrm{G}}
\newcommand\Gmeas{\G_{\rm m}}
\newcommand\vg{\mathbf{g}}
\renewcommand\vr{\mathbf{r}}
\newcommand\vv{\mathbf{v}}
\newcommand\vl{\boldsymbol{\ell}}
\newcommand\vk{\mathbf{k}}
\newcommand\vkeff{\vk_{\rm eff}}
\newcommand\keff{k_{\rm eff}}
\newcommand\bhat[1]{\mathbf{\hat{#1}}}
\newcommand\expect[1]{\left\langle #1\right\rangle}

\newcommand\Nmeas{N_{\rm m}}
\newcommand\temp{\mathcal{T}}
\newcommand\vOmega{\boldsymbol{\Omega}}

\newcommand\Rb{$^{87}$Rb}

\DeclareSIUnit{\eotvos}{E}

\begin{document}

\title{Tensor gravity gradiometry with a single-axis atom gradiometer}
\author{Ryan J. Thomas}\email{ryan.thomas@anu.edu.au}
\author{Samuel Legge}
\author{John D. Close}
\affiliation{Department of Quantum Science and Technology, The Australian National University, Canberra, Australia}

\begin{abstract}
	
We propose a method for using a single-axis atom interferometric gravity gradiometer to measure off-diagonal elements of the gravity gradient tensor.  By tilting the gradiometer, the measured gradient becomes a linear combination of different components of the gravity gradient tensor, and through multiple measurements at different tilts the separate tensor components can be inferred.  We present a theoretical and numerical investigation of this technique, both for terrestrial surveys where the tilt is statically set by the user and for surveys where a strapdown sensor is dynamically tilted by the motion of the platform.  We show that the gradiometer's sensitivity to the vertical gravity gradient is only slightly reduced by this method while allowing for more gradiometer information to be obtained.  Major sources of error and loss of sensitivity on dynamic platforms are shown to be mitigated using an optical-gimbal technique employing commercially-available fibre-optic gyroscopes and tip-tilt mirrors.

\end{abstract}

\maketitle

\section{Introduction}

Gravity gradiometry is a widespread geophysical surveying technique with applications in oil, gas, and mineral exploration \cite{bell_gravity_1998,pawlowski_gravity_1998,dransfield_pdf_2007,salem_interpretation_2013,veryaskin_gravity_2021}, bathymetry \cite{wan_sensitivity_2019,xu_predicting_2024}, civil engineering and underground structure detection \cite{edwards_gravity_1997,stray_quantum_2022,qiao_application_2025}, and geodesy \cite{rummel_goce_2011}.  Although gravimeters are often simpler, more compact devices, gradiometers are immune to platform vibrations, and gradiometer data is more sensitive to shorter spatial wavelengths, the combined effect of which is to improve detection and discrimination of gravitational anomalies \cite{dransfield_pdf_2007}, especially on mobile platforms.  Furthermore, measurements of the full gradient tensor, including on- and off-diagonal elements, provide rich information about subsurface properties which can aid in depth reconstruction \cite{salem_interpretation_2013,kebede_processing_2021}.  

Atom interferometers are a promising technology for gravity gradiometry \cite{snadden_measurement_1998}, having already been used for metrological \cite{fixler_atom_2007,rosi_precision_2014} and fundamental physics experiments \cite{asenbaum_phase_2017,asenbaum_atom-interferometric_2020,barrett_correlative_2015,weiner_flight_2020}.  Recent work aims to move these atom interferometric gravity gradiometers (AIGGs) outside of the laboratory \cite{janvier_compact_2022,lyu_compact_2022}, notably for the detection of underground structures \cite{stray_quantum_2022}, use in marine gravity gradient surveys \cite{delta_g_company_2025}, and gravitational map-matching for inertial navigation \cite{lellouch_integration_2025}. With few exceptions \cite{biedermann_testing_2015,stolzenberg_multi-axis_2025}, however, existing AIGGs are single-axis devices which only measure the vertical variation of the vertical component of gravity, denoted here as $\G_{zz} = \partial g_z/\partial z$.   The reason for this choice is twofold.  First, terrestrial gravity accelerates atoms in the vertical direction, so the longest interferometer time, and thus highest sensitivity, is achieved with a vertically oriented interferometer.  Second, maximum rejection of common-mode interferometer phase noise is achieved when the two atomic samples are interrogated with the same lasers and with only vacuum between the samples \cite{biedermann_testing_2015}.  Thus, the preferred configuration is to vertically separate the two atomic samples and to interrogate them with a vertically oriented laser.  While in micro-g environments, where there is no preferred direction, it is relatively trivial to measure the other diagonal elements of the gravity gradient tensor, $\G_{xx}$ and $\G_{yy}$ \cite{carraz_spaceborne_2014}, in terrestrial environments the need to maximize free-fall duration and reject common-mode phase noise makes it difficult to measure the off-diagonal elements.

In this article we propose a method for measuring off-diagonal elements of the gravity gradient tensor using a single-axis strap-down AIGG.  By tilting the gradiometer, either deliberately or through rotation of the platform, the measured gradient becomes a linear combination of different elements of $\G$.  Standard analysis techniques, together with auxiliary data from tilt sensors, can then be used to infer the individual tensor elements.  Dynamic effects, arising primarily from platform rotations, are shown to be correctable using commercially-available fibre-optic gyroscopes and tip-tilt mirrors.  We show that additional tensor elements can be obtained with minimal impact on the sensitivity to $\G_{zz}$.

\section{Gradiometer with a fixed tilt}
\label{sec:fixed-tilt}

\begin{figure}[t]
	\centering
	\includegraphics[width=\columnwidth]{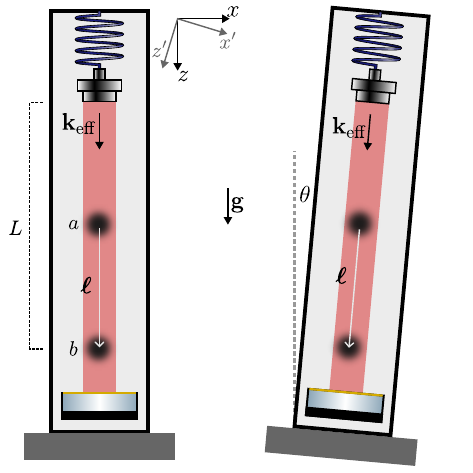}
	\caption{Simplified schematic of an AIGG, comprising two atomic samples, $a$ and $b$, that are separated by vector $\vl$.  Both samples participate in simultaneous Mach-Zehnder interferometers with pulse separation time $T$ and effective wavevector $\vkeff$.  The samples are launched upward to a height $h = \frac{1}{2}gT^2$ with the constraint $\ell + h = L$.  The AIGG can also be rotated about the $y$ axis by an angle $\theta$ with respect to the vertical (right).}
	\label{fg:schematic}
\end{figure}

We consider the single-axis AIGG depicted in Fig.~\ref{fg:schematic}, consisting of two atomic samples, labeled $a$ and $b$ and located at $\vr_a$ and $\vr_b$, which are separated by the vector $\vl = \vr_b - \vr_a$.  
For both samples we run simultaneous Mach-Zehnder interferometers with pulse separation time $T$, where the momentum states are coupled by a two-photon transition (either Raman or Bragg) with effective wavevector $\vkeff$.  At the end of the interferometer sequence, we measure the population of atoms in the initial state, denoted $p_j$ for $j \in \{a,b\}$, which can be written as \cite{duan_operating_2014},
\begin{equation}
	p_j = B_j + \frac{C_j}{2}\cos\Phi_j,
\end{equation}
for fringe offsets $B_j$, fringe contrasts $C_j$, and interferometer phases $\Phi_j$. Here, $\Phi_j = \vkeff\cdot\vg(\vr_j) T^2 + \varphi_j$ for local gravity $\vg(\vr_j)$ at $\vr_j$ and variable phase $\varphi_j$, which is comprised of user-defined laser phases, laser phase noise, and vibration noise. Ideally, $B_j = 1/2$ and $C_j = 1$, but in realistic atom interferometers we typically have a contrast $C_j < 1$ due to the nonzero spatial and velocity widths of the atomic sample. By interrogating both samples simultaneously, we can ensure that laser phase noise and vibration noise is common to both interferometers, which means that $\varphi_b - \varphi_a = 0$ even in the presence of significant vibrational and laser phase noise \cite{snadden_measurement_1998}.  Writing $\vg(\vr_j) = \vg + \G\vr_j$ for gravity gradient tensor $\G$, the differential phase $\Delta\Phi = \Phi_b - \Phi_a$ is then
\begin{equation}
	\Delta\Phi = \vkeff\cdot\G\vl T^2,
	\label{eq:differential-phase}
\end{equation}
which shows that we measure the variation in the $\vl$ direction of the $\vkeff$ component of $\mathbf{g}$, or $\G_{\vkeff,\vl}$.  Nominally, $\vkeff = \keff \bhat{z}$ and $\vl = \ell\bhat{z}$, so that $\Delta\Phi = \keff\ell T^2 \G_{zz}$.  In order to maximize the scale factor $\keff\ell T^2$ for a given device size $L$, we assume that both samples are launched upward to a height $h = \frac{1}{2}g T^2$ and that $h + \ell = L$.

\subsection{Gradiometry in two dimensions}
\label{ssec:2d}
To illustrate how a tilted single-axis AIGG can be used for tensor gradiometry, we first consider the simpler two-dimensional (2D) case where the only nonzero components of $\G$ are $\G_{zz}$, $\G_{xx}$, and $\G_{xz}$.  We assume that the AIGG is rotated about the $y$ axis by an angle $\theta$, so that in the rotated frame $\vkeff = \keff\bhat{z}'$ and $\vl = \ell\bhat{z}'$, while in the frame affixed to the earth $\vkeff = \keff(\cos\theta\bhat{z} + \sin\theta\bhat{x})$ and $\vl= \ell(\cos\theta\bhat{z} + \sin\theta\bhat{x})$.  Defining the measured gradient as $\Gmeas = \G_{z'z'} = \Delta\Phi/(\keff \ell T^2)$, we have
\begin{equation}
	\Gmeas(\theta) = \G_{zz}\cos 2\theta + \G_{xz}\sin 2\theta,
	\label{eq:measured-gradient-2d}
\end{equation}
where we have used the fact that $\G$ is symmetric and traceless in free space \footnote{Since the gravitational field $\mathbf{g}$ is the gradient of a potential $V(\vr)$, symmetry in $\G$ is due to the commutativity of partial derivatives. $\G$ is traceless in free space since $\nabla^2 V(\vr) = 0$.}.  Equation \eqref{eq:measured-gradient-2d} implies that if we make at least two measurements of the gradient at nondegenerate angles, then we can reconstruct both independent components of $\G$.  In practice, however, we would make multiple measurements over a range of $\theta$ and estimate $\G_{zz}$ and $\G_{xz}$ using a least-squares approach.  Supposing that we take $\Nmeas$ measurements, and that the measurement variance $\delta\Gmeas^2$ is $\theta$ independent, then the variances for the tensor components are
\begin{subequations}
	\begin{align}
		\delta\G_{xz}^2 &= \frac{\delta\Gmeas^2}{\Nmeas\expect{\sin^2 2\theta}} \approx \frac{\delta\Gmeas^2}{4\Nmeas\expect{\theta^2}},\label{eq:gxz-2d-variance}\\
		\delta\G_{zz}^2 &= \frac{\delta\Gmeas^2}{\Nmeas\expect{\cos^2 2\theta}} \approx \frac{\delta\Gmeas^2}{\Nmeas},\label{eq:gzz-2d-variance}
	\end{align}
\end{subequations}
when $\expect{\theta} = 0$.  As expected, the variance in $\G_{xz}$ decreases as the spread in tilts is increased, up to $\theta = \pm \pi/4$, while the variance in $\G_{zz}$ behaves in the opposite fashion.  For small angles, the variance in $\G_{zz}$ is largely unchanged, and the optimum measurement strategy when $\theta$ is set by the user is then to conduct multiple measurements at $\theta = \pm\theta_{\rm max}$ with $\expect{\theta} = 0$, which maximizes the variance of $\theta$, $\expect{\theta^2} = \theta_{\rm max}^2$, in the small-angle limit.  

Of course, we cannot assume that the measurement noise in an atom interferometer is independent of the angle, since the atoms now have an effective acceleration transverse to the lasers of $g\sin\theta$. Leaving a discussion of systematic errors to the end of this section, the major effect of the tilt on the gradiometer is a loss of sensitivity due to errors in the interferometer pulse areas. Let $\delta_n$ be the difference between the actual pulse area $\mathcal{A}_n$ and the nominal pulse area for pulses labeled by $n = \{0,1,2\}$.  In the limit $\delta_n \ll 1$, the fringe contrast becomes
\begin{align}
	C_j &\approx 1 - \frac{1}{2}\delta_0^2 - \frac{1}{4}\delta_1^2 - \frac{1}{2}\delta_2^2\label{eq:fringe-contrast}\\
	&\approx 1 - 128\pi^2\frac{h^4 \sin^4\theta}{w^4},
\end{align}
for a common Gaussian laser beam waist $w$.  Here, we have assumed that we can neglect the spatial size of the sample, a restriction we will later relax.  Assuming equal and uncorrelated interferometer phase variances $\delta\Phi^2$ for each individual interferometer, which are due to both atom shot noise and detection noise, the variance in the measured gradient is then
\begin{equation}
	\delta\Gmeas^2 = \frac{\delta\Phi^2 g^2}{2\Nmeas C(\theta,h)^2\keff^2 h^2\ell^2},
	\label{eq:gmeas-variance}
\end{equation}
where we have assumed that the fringe contrasts for each interferometer are the same, $C_1 = C_2 = C(\theta,h)$.  Straightforward minimization of the variance $\delta\G_{xz}^2$ in Eq.~\eqref{eq:gxz-2d-variance} yields an optimum tilt angle of $\theta_{\rm max}^4 = w^4/(640\pi^2 h^4)$, with a fringe contrast of $C = 4/5$ and resulting minimum variances of
\begin{subequations}
	\begin{align}
		\delta\G_{zz}^2 &= \frac{25}{32}\frac{\delta\Phi^2 g^2}{\Nmeas\keff^2 h^2\ell^2},\label{eq:gzz-2d-minimum-variance}\\
		\delta\G_{xz}^2 &= \sqrt{\frac{3125}{128}}\pi\frac{\delta\Phi^2 g^2}{\Nmeas \keff^2 w^2\ell^2},
		\label{eq:gxz-2d-minimum-variance}
	\end{align}
\end{subequations}
where $\delta\G_{xz}^2$ is independent of the launch height $h$ and thus $T$.  Therefore, we can choose $h$ (or $T$) based on the sensitivity that we want to attain in our measurement of $\G_{zz}$, which in turn determines the size of the AIGG, and there is no trade-off in the sensitivity in $\G_{xz}$ that we can reach.

To illustrate the design choices and resulting sensitivities, we consider an AIGG using \Rb{} atoms where $\keff = 4\pi/\lambda$ with $\lambda \approx \SI{780}{\nano\meter}$.  For the AIGG to be useful as a field instrument, its total length $L$ needs to be constrained; here, we will assume that $L = \SI{1}{\meter}$ \cite{lyu_compact_2022}.  To maximize sensitivity when $L = \ell + h$, we need $h = \ell = L/2 = \SI{0.5}{\meter}$, which yields $T = \SI{320}{\milli\second}$.  To minimize the variance in Eq.~\eqref{eq:gxz-2d-minimum-variance}, we further assume that $w = \SI{25}{\milli\meter}$, which is at the high end of beam waists used in atom interferometry \cite{asenbaum_phase_2017}.  Finally, since AIGGs perform a differential measurement, they can operate at \cite{janvier_compact_2022} or near \cite{asenbaum_phase_2017} the atom shot noise limit.  We therefore assume that $\delta\Phi = \SI{2}{\milli\radian}$, which corresponds to the atom shot noise limit for $2.5\times 10^5$ atoms.  The resulting sensitivities are then $\delta\G_{zz} \approx \SI{4.3}{\eotvos}$ and $\delta\G_{xz} \approx \SI{390}{\eotvos}$ per shot ($\SI{1}{\eotvos} = \SI{e-9}{\second^{-2}}$).

It is instructive to consider the minimum variances that we can achieve in alternative gradiometer configurations.  The simplest alternative is to alternate between $\theta = 0$ and $\theta = \pi/4$: the former only measures $\G_{zz}$, and the latter only measures $\G_{xz}$.  When the AIGG is tilted at $\theta = \pi/4$, we now find an optimum $h^4 = w^4/(160\pi^2)$ for measuring $\G_{xz}$, and we have minimum variances
\begin{subequations}
	\begin{align}
		\delta\G_{zz}^2 &= \frac{\delta\Phi^2 g^2}{\Nmeas\keff^2 h^2 \ell^2},\label{eq:gzz-2d-minimum-variance-45-deg}\\
		\delta\G_{xz}^2 &= \sqrt{\frac{3125}{8}}\pi\frac{\delta\Phi^2 g^2}{\Nmeas \keff^2 w^2\ell^2},
		\label{eq:gxz-2d-minimum-variance-45-deg}
	\end{align}
\end{subequations}
where we have assumed that we make $\Nmeas/2$ measurements at each of the two angles, which slightly increases the variance $\delta\G_{zz}^2$ relative to the small tilt case in Eq.~\eqref{eq:gzz-2d-minimum-variance}, to give $\delta\G_{zz} \approx \SI{5}{\eotvos}$ per shot.  The variance $\delta\G_{xz}^2$ is exactly four times as large as the variance for the small-angle case, Eq.~\eqref{eq:gxz-2d-minimum-variance}, or \SI{780}{\eotvos} per shot for the parameters given above.   

A more complicated alternative is to design an AIGG that can prepare two atomic samples where $\vl\perp\vkeff$, but where the two samples still interact with the same interferometer laser so that $\ell \sim w$.  The measured gradient in this case is equivalent to Eq.~\eqref{eq:measured-gradient-2d} with the replacement $\theta\to\theta + \pi/2$.  Assuming that $\ell = w$, with the samples symmetrically located on either side of the interferometer laser beam at $\pm w/2$, and assuming that we can use the full device size $h = L = \frac{1}{2} g T^2$ for the interferometer, we have minimum variances
\begin{subequations}
	\begin{align}
		\delta\G_{zz}^2 &= \frac{243\pi^2}{32}\frac{\delta\Phi^2 g^2}{\Nmeas \keff^2 w^4},\label{eq:gzz-2d-minimum-variance-perp}\\
		\delta\G_{xz}^2 &= \frac{9}{8}\frac{\delta\Phi^2 g^2}{\Nmeas \keff^2 w^2 L^2}\label{eq:gxz-2d-minimum-variance-perp},
	\end{align}
\end{subequations}
when we use the same technique of alternately tilting the gradiometer to $\pm\theta_{\rm max} = \pm w/(3\sqrt{3} \pi h)$.  Thus, for $L = \SI{1}{\meter}$, $\delta\G_{xz} \approx \SI{52}{\eotvos}$, and $\delta\G_{zz} \approx \SI{1.6e4}{\eotvos}$.  Although the sensitivity to $\G_{xz}$ for the $\vl\perp\vkeff$ gradiometer is better than for the $\vl\parallel\vkeff$ design, the sensitivity to $\G_{zz}$ is significantly worse due to the smaller separation between the samples, and in some applications, such as estimating source locations through the adaptive tilt method \cite{salem_interpretation_2013}, the large variance in $\G_{zz}$ will dominate the uncertainty.  Furthermore, for laser-cooled samples, the complexity of the $\vl\perp\vkeff$ design is higher than for the $\vl\parallel\vkeff$ design, and the expansion of the atomic samples during the interferometer may lead to those samples overlapping during detection.  Finally, errors due to wave-front aberrations may be larger in the $\vl\perp\vkeff$ configuration, since the two samples fall through different transverse parts of the beam, and this can lead to difficult-to-characterize systematic errors \cite{trimeche_active_2017,schkolnik_effect_2015}.  

\begin{figure}[tb]
	\centering
	\includegraphics[width=\columnwidth]{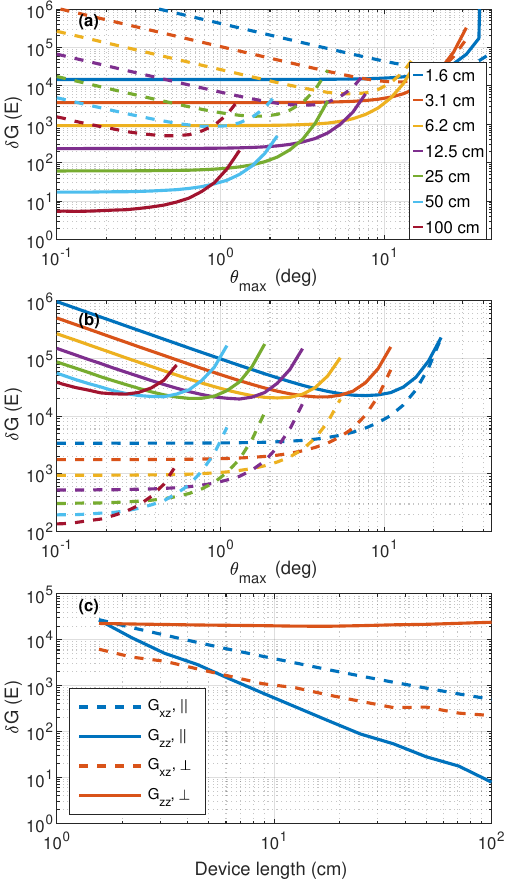}
	\caption{2D AIGG sensitivities when alternating the tilt of the gradiometer between $\pm\theta_{\rm max}$ for different total device lengths $L$ using a numerical model with temperature $\temp = \SI{3}{\micro\kelvin}$ and $w = \SI{25}{\milli\meter}$.  \textbf{(a)} Sensitivities $\delta\G_{zz}$ (solid lines) and $\delta\G_{xz}$ (dashed lines) for $\vl\parallel\vkeff$.  \textbf{(b)} Sensitivities $\delta\G_{zz}$ (solid lines) and $\delta\G_{xz}$ (dashed lines) for $\vl\perp\vkeff$.  \textbf{(c)} Sensitivities $\delta\G_{zz}$ and $\delta\G_{xz}$ for the parallel and perpendicular cases when $\theta_{\rm max}$ is chosen to optimize sensitivity to either $\G_{xz}$ (parallel) or $\G_{zz}$ (perpendicular).}
	\label{fg:static-2d-sensitivity}
\end{figure}
So far, we have ignored the temperature of the samples, but all AIGGs will use atomic samples with a nonzero temperature, and thus nonzero velocity spread, which will cause the atomic samples to increase in size during the interferometer sequence.  This will, in turn, lead to reductions in contrast due to both increased pulse area errors and velocity-dependent phase shifts \cite{roura_overcoming_2014}.  To evaluate the loss in contrast, and thus sensitivity, from the nonzero temperature we use a numerical model of the interferometer \cite{github}, where we evolve a sample of atoms with initial positions and velocities drawn from a Maxwell-Boltzmann distribution according to a discrete-time representation of the atomic trajectories and atom-light interaction.  We evaluate the sensitivities of both the $\vl\parallel\vkeff$ and $\vl\perp\vkeff$ designs, and these are shown in Fig.~\ref{fg:static-2d-sensitivity}.  As expected from Eq.~\eqref{eq:gxz-2d-minimum-variance}, Fig.~\ref{fg:static-2d-sensitivity}a shows that for large $L$ there is an optimum $\theta_{\rm max}$ which minimizes $\delta\G_{xz}$ and for which there is only a small increase in $\delta\G_{zz}$.  The sensitivity $\delta\G_{xz}$ also decreases with increasing $L$, since the separation between the samples $\ell = L/2$ also increases with the device size.  In contrast, Fig.~\ref{fg:static-2d-sensitivity}b shows that while $\delta\G_{xz}$ decreases with increasing $L$ for the $\vl\perp\vkeff$ design, the minimum $\delta\G_{zz}$ is independent of the device length, as expected from Eq.~\eqref{eq:gzz-2d-minimum-variance-perp}.  For both gradiometer configurations, the minimum sensitivities are not as small as predicted from the analytic theory, since the finite temperature of the samples reduces the fringe contrast.  In particular, for the $\vl\perp\vkeff$ design, the minimum sensitivity deviates more than for the $\vl\parallel\vkeff$ design, since the atomic samples are off-center from the beam and thus see larger variations in the laser power across the sample.  Figure~\ref{fg:static-2d-sensitivity}c compares the sensitivities for the optimum choice of $\theta_{\rm max}$ between the two different designs.  For $\vl\parallel\vkeff$, the sensitivities improve with increasing device size $L$ as $\delta\G_{xz} \propto L^{-1}$ and $\delta\G_{zz} \propto L^{-2}$, whereas for $\vl\perp\vkeff$ $\delta\G_{xz}\propto L^{-1}$ and $\delta\G_{zz}$ is approximately constant.  At large device lengths, the increased spatial size of the sample combined with the variation in laser intensities away from the beam center leads to only small differences in the optimum sensitivities in $\G_{xz}$ for the two different design choices.

Finally, we return to the question of systematic errors induced by the tilt of the gradiometer. Although each individual atom interferometer may acquire several systematic errors related to the motion of the atoms transverse to the lasers, such as errors from differential AC Stark shifts, two-photon light shifts \cite{gauguet_off-resonant_2008}, Coriolis accelerations \cite{lyu_compact_2022}, and wave-front aberrations \cite{schkolnik_effect_2015,trimeche_active_2017,hu_analysis_2017}, large interferometer beam waists and careful design of the gradiometer will ensure that those contributions to the differential phase $\Delta\Phi$ will be negligible. In addition to sources of systematic error already characterized for AIGGs \cite{lyu_compact_2022}, the major new sources of systematic error will be related to noise or bias in the measurement of the tilt and false gravity gradients induced by the combination of tilt and differential launch velocities. With regards to the former, for small tilt error (either random noise or bias) $\delta\theta_t \ll 1$, the bias in the measured gradient is
\begin{equation}
	\expect{\delta\Gmeas} \approx 2\expect{\delta\theta_t}\G_{xz} - 2\expect{\delta\theta_t^2}\Gmeas.
\end{equation}
For geophysical signals where $\Gmeas \approx \SI{3000}{\eotvos}$ and $\G_{xz} \approx \SI{10}{\eotvos}$, we would require a bias or noise in the tilt measurement of at least \SI{10}{\milli\radian} to reach a systematic error of $\expect{\delta\Gmeas} = \SI{1}{\eotvos}$. High-quality tilt sensors routinely have biases and noise below $\SI{1}{\milli\radian}$, and thus systematic errors from tilt biases are likely negligible.

Differential Coriolis accelerations, caused by a difference in launch velocity, may have a much larger impact on the gradient measurement. For an Earth rotation rate of $\Omega_E \approx \SI{72}{\micro\radian/\second}$, the measured gradient is
\begin{equation}
	\Gmeas = \G_{zz}\cos 2\theta + \G_{xz}\sin 2\theta + 2\Omega_E \Delta v_L\sin\theta,
\end{equation}
where $\Delta v_L$ is the difference in launch velocity between the two atomic samples. For $\G_{xz} = \SI{10}{\eotvos}$, we will have a systematic error of \SI{1}{\eotvos} if $\Delta v_L \approx \SI{15}{\micro\meter/\second}$, compared to a launch velocity of approximately \SI{3}{\meter/\second} needed to reach a total height of $h = \SI{0.5}{\meter}$. Previous methods for eliminating the effect of differential launch velocities by rotating the apparatus $180^\circ$ do not apply for tensor measurements, since this is equivalent to switching $\theta\to -\theta$ and is the method by which we measure $\G_{xz}$. Differential launch velocities can potentially be calibrated in a laboratory by using large tilt angles where $\sin 2\theta \not\approx 2\sin\theta$, so that an appropriate correction can be applied when the AIGG is transported to and used in the field.

\subsection{Gradiometry in three dimensions}
\label{ssec:3d}
We now consider tilted gradiometer measurements in three dimensions.  We again assume that $\vl\parallel\vkeff$, except now $\vkeff = \keff(\sin\theta\cos\phi\bhat{x} + \sin\theta\sin\phi\bhat{y} + \cos\theta\bhat{z})$.  Using $\G_{xx} + \G_{yy} + \G_{zz} = 0$, the measured gradient is
\begin{align}
	\Gmeas &= \G_{zz}(\cos^2\theta - \sin^2\theta\sin^2\phi) + \G_{xx}\sin^2\theta\cos 2\phi\notag\\
	&\quad + (\G_{xz}\cos\phi + \G_{yz}\sin\phi)\sin 2\theta\notag\\
	&\quad + \G_{xy}\sin^2\theta\sin 2\phi,
	\label{eq:measured-gradient-3d}
\end{align}
and, for a measurement along $\phi = 0$ with $\theta \ll 1$, reduces to
\begin{equation}
	\Gmeas \approx \G_{zz} + 2\theta\G_{xz} + \theta^2\left(\G_{xx} - \frac{1}{2}\G_{zz}\right).
	\label{eq:measured-gradient-3d-phi=0}
\end{equation}
Equation \eqref{eq:measured-gradient-3d-phi=0} is different from Eq.~\eqref{eq:measured-gradient-2d} by the addition of a $\G_{xx}\theta^2$ term, which makes inferring the components of $\G$ from a set of measurements at different $\theta$ more complicated.  First, the extra parameter $\G_{xx}$ means that we can no longer take measurements only at $\pm\theta_{\rm max}$; the optimal strategy is now to take some measurements at $\theta = 0$ and the rest evenly split between $\pm\theta_{\rm max}$, with the exact ratio depending on the relative measurement variances at these two angles.  Second, the variances in both $\G_{zz}$ and $\G_{xz}$ increase both from the extra degree of freedom and the change in the distribution of $\theta$ values.  Finally, the variance in $\G_{xx}$ estimated using a least-squares fit to Eq.~\eqref{eq:measured-gradient-3d-phi=0} is large and scales as $g^4T^4/w^4$, implying that at long interferometer times, where $\G_{zz}$ sensitivity is maximized, the sensitivity to $\G_{xx}$ is minimized.  

Similar problems occur when estimating any of the horizontal gradient elements $\G_{xx}$, $\G_{yy}$, and $\G_{xy}$ from small tilts about the vertical, implying that we can only measure vertical elements $\G_{\{x,y,z\},z}$ precisely.  However, atom interferometers can be straightforwardly constructed so that they can operate at any angle \cite{darmagnac_de_castanet_atom_2024}, so the gradiometer can be rotated to $\theta = \pi/2$ if mounted on an appropriate rotation stage.  This suggests the following protocol for measuring the full gravity gradient tensor using a single-axis AIGG.  First, choose a total device length $L$ to attain a given sensitivity for $\G_{zz}$, which then fixes the optimum sensitivities for $\G_{xz}$ and $\G_{yz}$.  For large $L$ (long $T$), the optimum tilt angles for measuring $\G_{xz}$ and $\G_{yz}$ are then small angles.  For our reference parameters, with $L = \SI{1}{\meter}$ and $w = \SI{25}{\milli\meter}$, the optimum angle is $\theta_{\rm max} \approx \SI{10}{\milli\radian}$.  Given that the measurement noise in this situation is $\delta\Gmeas \approx \SI{4.3}{\eotvos}$ per shot and that typical gravity gradients from geophysical targets are $\mathord{\sim}\SI{50}{\eotvos}$, the contribution of the quadratic term in Eq.~\eqref{eq:measured-gradient-3d-phi=0} will be a factor of $\mathord{\sim}10^3$ below the measurement noise level, which is a negligible contribution and can be ignored \footnote{The contribution of the Earth's background gradient of $\mathord{\approx}\,\SI{3000}{\eotvos}$ can be subtracted from the measurement, leaving only signals from gravitational anomalies}.  Thus, our strategy is to first measure the vertical components of the gradient by taking measurements at $\pm\theta_{\rm max}$ at both $\phi = 0$ and $\phi = \pi/2$.  We then rotate the gradiometer to $\theta = \pi/2$ and take measurements at $\phi = 0$ and $\phi = \pi/4$, where we neglect $\phi = \pi/2$ because we can compute $\G_{yy}$ from $\G_{xx}$ and $\G_{zz}$.  The optimum interferometer time for $\theta = \pi/2$ is $T_{\rm opt}^8 = w^4/(24\pi^2 g^4)$ for a minimum variance in $\G_{xx}$ of
\begin{equation}
	\delta\G_{xx}^2 = \sqrt{486}\pi \frac{\delta\Phi^2 g^2}{\Nmeas \keff^2 w^2 \ell^2},
	\label{eq:gxx-minimum-variance-3d}
\end{equation}
or $\delta\G_{xx} \approx \SI{816}{\eotvos}$ per shot, which applies to all horizontal components.  

As with the 2D case, it is instructive to consider a design where $\vl\perp\vkeff$.  Assuming that $\vkeff$ is as above, and $\vl$ is rotated about the $y$ axis by $+\pi/2$, the measured gradient is
\begin{align}
	\Gmeas &= \frac{1}{2}\big(\G_{xx}[1 + \cos^2\phi] + \G_{yy}[1 + \sin^2\phi]\notag\\
	&\quad + \G_{xy}\sin 2\phi\big)\sin 2\theta\notag\\
	&\quad + \big(\G_{xz}\cos\phi + \G_{yz}\sin\phi\big)\cos 2\theta.
	\label{eq:measured-gradient-3d-perp}
\end{align}
In contrast to Eq.~\eqref{eq:measured-gradient-3d}, no components of $\G$ have leading-order terms $\theta^2$ when $\theta \ll 1$, which means that we can use the technique of alternating between small angles $\pm\theta_{\rm max}$ without having to rotate the entire apparatus by $\pi/2$.  The measurement protocol would then be to conduct measurements at $\pm\theta_{\rm max}$ at angles $\phi = \{0,\pi/4,\pi/2\}$.  Based on the analysis in Sec.~\ref{ssec:2d}, the sensitivity for measuring $\G_{xx}$ is the same as the sensitivity in measuring $\G_{zz}$, which is $\delta\G_{xx} = \SI{1.6e4}{\eotvos}$ and is limited by the beam waist $w$.

\section{Gradiometry on a dynamic platform}
\label{sec:dynamic}

\subsection{Contrast reduction}

When used for gravity surveying or inertial navigation, gravity gradiometers are deployed on moving platforms (so-called ``moving-base'' gradiometers) where the platform dynamics can impact the gradient measurements \cite{bell_gravity_1998,chan_superconducting_1987}.  Unlike gravimeters, which are strongly affected by platform vibrations, the high common-mode rejection of gradiometers means that they are essentially immune to platform vibrations, making them ideal for marine or aerial surveys.  Gradiometers are highly sensitive to rotations, however, where the centrifugal force leads to an effective gravity gradient tensor $\G_{\rm cf} = \vOmega\vOmega^\mathrm{T}$ \cite{chan_superconducting_1987} for angular velocity vector $\vOmega$, which means that gradiometers must be gyroscopically stabilized with residual rotation rates of $<\SI{30}{\micro\radian/\second}$ to achieve $\SI{1}{\eotvos}$ accuracy.

In contrast to classical gradiometers, which require bulky mechanical gimbals for rotational stabilization, AIGGs can be employed in a strap-down configuration and the rotation stabilized using an ``optical gimbal.''  In this design, rather than gyroscopically stabilizing the entire AIGG, an optical gimbal stabilizes only $\vkeff$.  To understand this technique, we first calculate the interferometer phase without optical gimbaling in the frame affixed to the earth where the trajectory of the atoms is $\vr(t) = \vr_0 + \vv_0 t + \frac{1}{2}\vg t^2$ for initial positions $\vr_0$ and velocities $\vv_0$, and where $\vg$ is assumed to be evaluated at $\vr_0$.  Assuming that the AIGG rotates about the $y$ axis with rotation rate $\Omega(t) = \Omega_0 + \dot{\Omega}_0 t$, where $\dot{\Omega}_0$ is the angular acceleration, in the inertial frame the wavevector is now time dependent: $\vkeff(t) = \keff(\cos\theta(t) \bhat{z} + \sin\theta(t) \bhat{x})$, where $\theta(t) = \int \Omega(t') dt' = \theta_0 + \Omega_0 t + \frac{1}{2}\dot{\Omega}_0 t^2 = \theta_0 + \delta\theta(t)$ with $\delta\theta(t) \ll 1$.  For instantaneous light pulses occurring at $t_n = \{-T,0,T\}$, the total interferometer phase for a single atom, $\Phi = \vkeff(T)\cdot\vr(T) - 2\vkeff(0)\cdot\vr(0) + \vkeff(-T)\cdot\vr(-T)$, is
\begin{align}
	\Phi &= \keff T^2(g_z' - \Omega_0^2 z_0' + 2v_{x0}'\Omega_0)\notag\\
	&\quad - 3\keff T^3(\dot{\Omega}_0\Omega_0 z_0' + \Omega_0^2 v_{z0}' - \dot{\Omega}_0 v_{x0}'),
	\label{eq:rotation-single-atom-phase-uncorrected}
\end{align}
where primed coordinates are in the frame rotated by $\theta_0$ (see Fig.~\ref{fg:schematic}), and we have kept terms only up to $T^3$.  Similarly to classical gradiometers, the centrifugal acceleration $\Omega_0^2 z_0'$ leads to an effective gravity gradient $\Omega_0^2$.  Unlike classical gradiometers, however, the combination of the Coriolis acceleration $2v_{x0}'\Omega_0$ and the Maxwell-Boltzmann distribution in initial velocities $(v_{x0}',v_{z0}')$ leads to a rapid loss in interferometer contrast and thus sensitivity. For velocity standard deviations $(\sigma_{x'},\sigma_{z'})$, the leading-order contribution to the interferometer contrast is $C \propto \exp[-2\keff^2\sigma_{x'}^2 \Omega_0^2 T^4]$ \cite{darmagnac_de_castanet_atom_2024}, which drops by a factor of $1/e$ for $\Omega_0 = \SI{25}{\micro\radian/\second}$ when using a sample of \Rb{} atoms at a temperature of $\mathcal{T} = \SI{3}{\micro\kelvin}$ and $T = \SI{320}{\milli\second}$. Therefore, for the long interferometer times necessary for high-sensitivity gravity gradiometry, a method for preserving fringe contrast is needed. Although there are interferometer topologies that reduce or eliminate the effect of Coriolis accelerations on the interferometer phase, these have significant trade-offs either in terms of sensitivity to gravity gradients \cite{marzlin_state_1996} or sensitivity to vibrations \cite{mcguirk_sensitive_2002,perrin_proof--principle_2019}. Since insensitivity to platform vibration is a key advantage of gradiometers over gravimeters, we do not consider these interferometer topologies here.


The standard method to compensate for this loss in fringe contrast is to use a tip-tilt system to rotate the retroreflecting mirror by an angle $\theta_m(t) = \delta\theta(t)$ during the interferometer sequence \cite{dickerson_multiaxis_2013,sugarbaker_enhanced_2013,darmagnac_de_castanet_atom_2024}.  This yields an effective wavevector $\vkeff(t) = \keff\cos\theta_m(t)(\cos\theta_0 \bhat{z} + \sin\theta_0 \bhat{x})$, where the direction of the wavevector remains constant but its magnitude changes.  This method works well for small rotation rates, such as at Earth's rotation rate \cite{sugarbaker_enhanced_2013,freier_atom_2017}, or for short interferometer times \cite{darmagnac_de_castanet_atom_2024}, but for the long interferometer times needed for high-sensitivity gradiometry the fringe contrast varies as $C \propto \exp[-\frac{9}{2}\keff^2 T^6(\sigma_{x'}^2\dot{\Omega}_0^2 + \sigma_{z'}^2\Omega_0^4)]$, with a $1/e$ angular velocity of $\Omega_0 = \SI{7}{\milli\radian/\second}$ or angular acceleration of $\dot{\Omega}_0 \approx \SI{53}{\micro\radian/\second^2}$ at $T = \SI{320}{\milli\second}$ and $\mathcal{T} = \SI{3}{\micro\kelvin}$.  The decay of the fringe contrast is now driven by the shortening of $\vkeff$.

There are two ways to fix the problem of a reduced $\keff$.  One method is to increase $\keff$ by increasing the average laser frequency \cite{asenbaum_atom-interferometric_2020} so that $\keff \to \keff\sec\theta_m(t)$, which has the advantage that for some laser systems this can be achieved with a simple change to a radio-frequency modulation \cite{templier_carrier-suppressed_2021}.  However, this technique has the drawback that significant changes to the laser frequency and power are required even for small angles.  For example, at $\delta\theta(2T) = \SI{5}{\milli\radian}$ the laser frequency for a \Rb{} interferometer would need to increase by $\mathord{\sim}\SI{2.5}{\giga\hertz}$, and in order to maintain a constant Rabi frequency, the optical powers would need to increase by $\mathord{\sim}\SI{1}{\watt}$.  The second method is to employ another tip-tilt mirror for the output coupler of the lasers to keep $\vkeff$ static.  If the output coupler angle changes by $\theta_c(t)$, then $\vkeff(t) = \keff\cos(\theta_c(t) - \theta_m(t))[\cos(\theta(t) - \theta_m(t))\bhat{z} + \sin(\theta(t) - \theta_m(t))\bhat{x}]$, where ideally, $\theta_m(t) = \theta_c(t) = \delta\theta(t)$, so that neither the direction nor magnitude of $\vkeff$ changes in the inertial frame.  In contrast to changing the laser frequency, this method does not require a frequency-agile laser nor significant reserve optical power, but adding a second tip-tilt mirror may instead present additional mechanical complications and possible redesigns of the optical delivery system in existing AIGGs.

Although ideally the tip-tilt mirrors or the laser frequency exactly track the platform rotation, in practice there will be some mismatch between these actuators and the actual platform.  Here, we only consider scale factor errors, most likely arising from tip-tilt mirror calibrations \cite{darmagnac_de_castanet_atom_2024}, since bias errors and angle-random walk in commercial navigation-grade fibre-optic gyroscopes contribute negligibly over the interferometer sequence.  We consider here only the case where there are two tip-tilt mirrors, but a similar analysis applies to the case of changing $\keff$.  We assume that $\theta_m(t) = [1 + \epsilon_m]\delta\theta(t)$ and $\theta_c(t) = [1 + \epsilon_m + \delta\epsilon]\delta\theta(t)$, where $\epsilon_m$ is the scale factor error for the retroreflection mirror, and $\delta\epsilon$ is the difference in scale factor errors between the mirror and output coupler.  Computing the interferometer phase to leading order in $T$ and $(\epsilon_m^2,\delta\epsilon^2)$, the fringe contrast is
\begin{equation}
	C(\Omega) = C_0\exp\Big[-\frac{\epsilon_m^2}{2}\keff^2\sigma_{x'}^2 T^4(2\Omega_0 + 3\dot{\Omega}_0 T)^2\Big],
	\label{eq:rotation-fringe-contrast-corrected}
\end{equation}
which decreases due to the residual Coriolis acceleration in the $x$ direction.  The phase difference between the two interferometers is then
\begin{equation}
	\Delta\Phi = \keff\ell T^2\Big(\G_{z'z'} - [\epsilon_m^2 + \delta\epsilon^2]\Omega_0^2\Big).
	\label{eq:rotation-differential-phase}
\end{equation}
For both Eqs.~\eqref{eq:rotation-fringe-contrast-corrected} and \eqref{eq:rotation-differential-phase}, the situation without optical gimbaling corresponds to $\epsilon_m = -1$ and $\delta\epsilon = 0$.  
 
\begin{figure}[tb]
	\centering
	\includegraphics[width=\columnwidth]{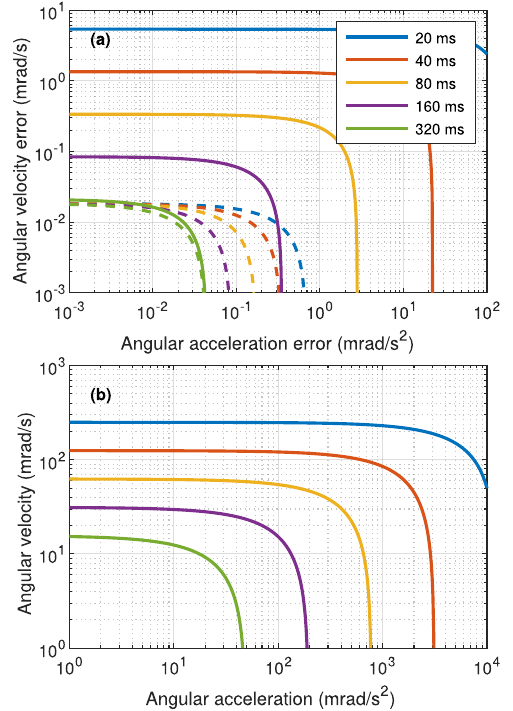}
	\caption{\textbf{(a)} Contours of the mirror's angular velocity and acceleration error which lead to a reduction in fringe contrast by a factor of $2$ (solid lines) and an error in the differential phase of \SI{1}{\eotvos} (dashed lines).  \textbf{(b)} Contours of the platform's angular velocity and acceleration which lead to a total mirror angle adjustment of \SI{10}{\milli\radian} and/or a required frequency shift of \SI{10}{\giga\hertz} for a laser frequency of \SI{384.230}{\tera\hertz} corresponding to the \Rb{} D2 transition.}
	\label{fg:rotation-contrast}
\end{figure}
In Fig.~\ref{fg:rotation-contrast}a we plot contours of where the fringe contrast has dropped to $C_0/2$ as a function of the error in the mirror's angular velocity and acceleration $(\epsilon_m\Omega_0,\epsilon_m\dot{\Omega}_0)$, and where we have assumed that $\delta\epsilon^2 = 2\epsilon_m^2$.  We also plot contours where the error in the measurement of $\G_{z'z'}$ is \SI{1}{\eotvos}.  Without correction, the level of rotational motion that can be tolerated is very small and is limited by the error in $\G_{z'z'}$ rather than the reduction in contrast at all interferometer times due to the residual centrifugal acceleration (RCA) $\delta\G_{\rm RCA} = (\epsilon_m^2 + \delta\epsilon^2)\Omega_0^2$.  However, depending on the scale factor error, the tolerable rotational motion can be significantly larger.  Although commercial fibre-optic gyroscopes can achieve scale factor errors of $\epsilon_m = 10^{-4}$, it is likely that the scale factor error will be dominated by the calibration of the tip-tilt mounts rather than the gyroscopes.  Recent work has shown that the scale factor error between gyroscopes and tip-tilt mirrors can be characterized at the $10^{-3}$ level \cite{darmagnac_de_castanet_atom_2024}, implying that $\Omega_0 < \SI{20}{\milli\radian/\second}$ and $\dot{\Omega}_0 < \SI{40}{\milli\radian/\second^2}$.  Furthermore, tip-tilt mirrors have a limited compliance range, typically on the order of $\theta_{\rm tt,max} = \SI{10}{\milli\radian}$, which in turn limits the allowable rotational motion.  Figure~\ref{fg:rotation-contrast}b plots contours of where the tip-tilt mirror(s) move by \SI{10}{\milli\radian} or the laser frequency shifts by \SI{10}{\giga\hertz}, and for these values $\Omega_0 < \SI{15}{\milli\radian/\second}$ and $\dot{\Omega}_0 < \SI{45}{\milli\radian/\second^2}$ for $T = \SI{320}{\milli\second}$, which correspond well with the allowable rates when considering likely scale factor errors.

\subsection{Tensor gradiometry on a dynamic platform}
Since contrast reduction, gradient errors, and accumulated tip-tilt mirror angles are primarily driven by the angular velocity of the platform, the clear strategy is to take measurements with the AIGG when $\Omega_0$ is minimized.  For platforms undergoing reasonably smooth oscillatory motion, such as boats and airplanes in relatively calm conditions, the times when $\Omega_0 \approx 0$ correspond to times when the tilt angle is maximized, which conveniently allows for measurements of off-diagonal tensor elements of $\G$.  Other tilt angles can be used if the accumulated rotation angle during the interferometer sequence is less than the tip-tilt mirror's compliance and if the expected gradient error $\delta\G_{\rm RCA}$ is below some threshold value $\delta\G_{\rm RCA,max}$.

\begin{figure}[t]
	\centering
	\includegraphics[width=\columnwidth]{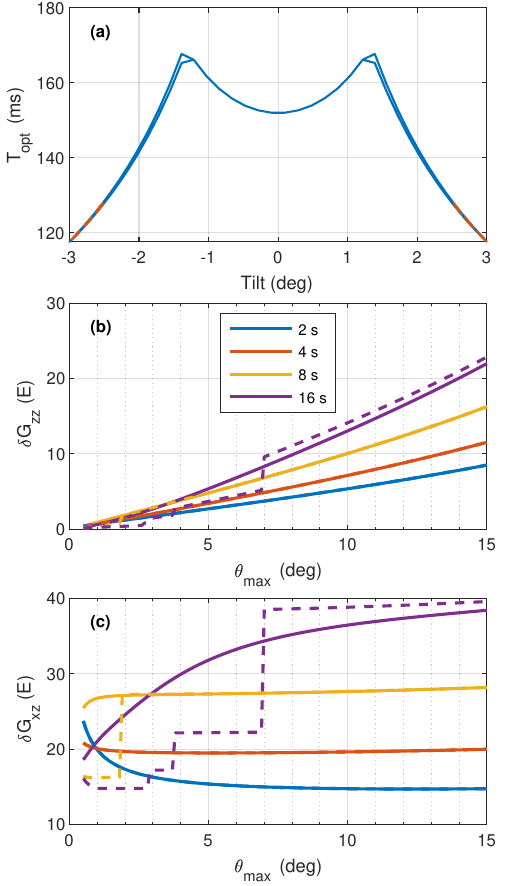}
	\caption{Performance of an AIGG on a dynamic platform with $L = \SI{1}{\meter}$, $\ell = \SI{0.5}{\meter}$, $\epsilon_m = 10^{-3}$, $\delta\epsilon^2 = 2\epsilon_m^2$, $w = \SI{25}{\milli\meter}$, $\theta_{\textrm{tt,max}} = \SI{10}{\milli\radian}$, and $\delta\Phi = \SI{2}{\milli\radian}$.  \textbf{(a)} Optimal interferometer time as a function of angle for $\theta_{\rm max} = 3^\circ$ and $2\pi/\omega = \SI{10}{\second}$ when measurements with $\delta\G_{\rm RCA} > \delta\G_{\rm RCA,max}$ are included (solid blue line) and when they are discarded (red dashed line).  \textbf{(b-c)} Uncertainties $\delta\G_{zz}$ and $\delta\G_{xz}$ after \SI{10}{\minute} of measurements for different maximum tilt angles $\theta_{\rm max}$ and angular periods $2\pi/\omega$ with $T_{\rm rep} = \SI{1}{\second}$.  Solid lines are when measurements with $\delta\G_{\rm RCA} > \delta\G_{\rm RCA,max}$ are included, and dashed lines are when they are discarded.}
	\label{fg:gradient-dynamic-platform}
\end{figure}
A possible measurement protocol is as follows.  We suppose that the tilt angle of the platform oscillates as $\theta(t) = \theta_{\rm max}\cos\omega t$ for maximum tilt $\theta_{\rm max}$ and angular frequency $\omega$; then, $\Omega_0 = -\omega\theta_{\rm max}\sin\omega t$ and $\dot{\Omega}_0 = -\omega^2\theta_{\rm max}\cos\omega t$.  Given a minimum repetition time $T_{\rm rep}$ for the AIGG, we conduct measurements at $N = 2\lfloor \pi/(\omega T_{\rm rep})\rfloor - 1$ equidistant times $t$, starting with $t = 0$, which ensures that we measure at $\pm\theta_{\rm max}$ and that the time between measurements is at least $T_{\rm rep}$.  For each measurement time corresponding to a set of $(\theta,\Omega_0,\dot{\Omega}_0)$, we compute the optimal interferometer time $T_{\rm opt}$ based on the AIGG parameters such as device length $L$, sample separation $\ell$, beam waist $w$, and scale factor errors $(\epsilon_m,\delta\epsilon)$.  $T_{\rm opt}$ is additionally subject to the constraints that $\delta\theta(2T_{\rm opt})$ is less than the tip-tilt mirror compliance and $T_{\rm opt}$ is less than the maximum interferometer time set by $T_{\rm max}^2 = 2(L - \ell)/g$.  Inferring $\Delta\Phi$ from population measurements $(p_a,p_b)$ requires multiple measurements at different common-mode phases, typically provided through random vibrational motion of the retroreflecting mirror, which thus requires multiple measurements at each $\theta$.  Once sufficient data has been collected, $\Delta\Phi$ can be inferred \cite{foster_method_2002,barrett_correlative_2015} and $\G_{z'z'}(\theta)$ can be calculated.  Depending on how consistent $\Omega_0$ is at different population measurements, a correction for the residual centrifugal acceleration can be applied to $\G_{z'z'}(\theta)$ with a corresponding increase in its variance by $4\epsilon_m^4\Omega_0^4$.  Alternatively, measurements where $\epsilon_m^2\Omega_0^2$ is greater than a threshold gradient error can be discarded.

In Fig.~\ref{fg:gradient-dynamic-platform}a we show the optimal interferometer time $T_{\rm opt}$ assuming that $\theta_{\rm max} = 3^\circ$ and $2\pi/\omega = \SI{10}{\second}$, both when measurements where $\delta\G_{\rm RCA} > \delta\G_{\rm RCA,max}$ are included and where such measurements are discarded.  When these measurements are included, we add an additional measurement variance corresponding to an uncertainty in $\epsilon_m$ and $\delta\epsilon$ equal to their values and a $10\%$ uncertainty in $\Omega_0$.  The multivalued nature of $T_{\rm opt}$ as a function of $\theta$ arises from there being two different values of $(\Omega_0,\dot{\Omega}_0)$ for each $\theta$, and the nonzero $\epsilon_m$ leads to different values of $T_{\rm opt}$.  The abrupt change in behavior for $|\theta| < 1.5^\circ$ is the effect of limited tip-tilt compliance, so shorter values of $T$ are required to keep $\delta\theta(2T_{\rm opt}) < \theta_{\rm tt,max}$.  When we discard measurements where $\delta\G_{\rm RCA} > \delta\G_{\rm RCA,max}$ (red dashed line), there is only a small range of angles where we can collect data.

Figures \ref{fg:gradient-dynamic-platform}b and c illustrate the sensitivities that can be achieved using the above protocol for different values of $\theta_{\rm max}$ and $\omega$ over a total measurement time of \SI{10}{\minute} when the minimum repetition time of the AIGG is $T_{\rm rep} = \SI{1}{\second}$.  The discontinuities for the cases when we discard measurements where $\delta\G_{\rm RCA} > \delta\G_{\rm RCA,max}$ (dashed lines) arise from the change in the number of measurements taken per cycle.  As $\theta_{\rm max}$ increases, $\Omega_0$ increases for angles away from the extremes $\pm\theta_{\rm max}$, and we discard more measurements, which leads to an increase in the measured gradient's variance.  For larger values of $\theta_{\rm max}$, worse sensitivities are seen for longer angular periods, and this is due to the condition that the total measurement duration is fixed at \SI{10}{\minute} so that fewer measurements are taken overall compared to shorter angular periods.  Generally speaking, we can see that for most $(\theta_{\rm max},\omega)$ the best strategy is to discard measurements where $\delta\G_{\rm RCA} > \delta\G_{\rm RCA,max}$, except for large maximum tilts and the slowest angular period.

\section{Conclusion}
\label{sec:conclusion}

In this paper, we have presented a simple method for using a tilted single-axis AIGG to measure off-diagonal elements of the gravity gradient tensor.  This technique is applicable both to terrestrial surveys where the tilt of the device is set by the user and for surveys on dynamic platforms where the tilt is governed by platform dynamics.  In the latter case, we showed that commercially available gyroscopes and tip-tilt mirrors can optically gimbal the gradiometer well enough that the motion of the platform has a negligible impact on the sensitivity of the measurement.

For the parameters we have considered, an AIGG using our tilted-axis tensor gradiometry technique has similar sensitivity to $\G_{zz}$ as existing commercial full-tensor gravity gradiometers but roughly a factor of $100$ worse sensitivity for the off-diagonal $(\G_{xz},\G_{yz})$ and horizontal $(\G_{xx},\G_{yy},\G_{xy})$ tensor elements \cite{difrancesco_gravity_2009}.  This reduction in sensitivity is a consequence of the atom shot-noise and the limited interrogation time due to the atoms freely-falling in a large gravitational field.  Nevertheless, given that gravity gradient anomalies in both geophysical \cite{bell_gravity_1998} and civil engineering \cite{stray_quantum_2022} contexts are on the order of $10-\SI{100}{\eotvos}$, our tensor gradiometry scheme can measure these off-diagonal signals with modest averaging times.

One avenue for improving the AIGG sensitivity without increasing the device size is to use large momentum transfer (LMT) techniques which increase $\keff$ \cite{giltner_atom_1995} and decrease the contribution of atom shot-noise, with some demonstrations showing an increase in $\keff$ by a factor of $100$ \cite{chiow_102ensuremathhbark_2011,mcdonald_80ensuremathhbark_2013,rodzinka_optimal_2024}. Increases to $\keff$ will be most useful for terrestrial applications where the device does not rotate, since the loss of fringe contrast due to rotations scales exponentially with $\keff^2$.  For dynamic platforms, a combination of reduced sample temperature (such as using Bose-Einstein condensates), improved gyroscopes, and improved tip-tilt mirror characterization will be necessary to take advantage of large momentum transfer techniques. As an alternative to increasing the sensitivity to off-diagonal elements, LMT can be used to reduce the required beam waist size [see Eq.~\eqref{eq:gxz-2d-minimum-variance}] and thus reduce laser power requirements. Care must be used, however, since a naive approach to LMT typically requires significantly more power \cite{szigeti_why_2012} than for standard momentum transfer. Recent developments in using machine learning to optimize interferometer pulses \cite{saywell_enhancing_2023}, including for LMT, will be useful in these contexts. Such techniques have the additional advantage that they configure the interferometer so that it is much less sensitive to changes in the laser intensity, which will increase the sensitivity of the off-diagonal measurements.

Improvements in the sensitivity to the off-diagonal and horizontal tensor elements can also be achieved by using multi-axis interferometers \cite{stolzenberg_multi-axis_2025}, but these require large baselines in both the horizontal and vertical direction and thus add substantial size, weight, and complexity.  Our tilted-axis method minimizes the required size of the apparatus, reduces design complexity, and can be employed with existing single-axis AIGGs with only minor modifications.

\begin{acknowledgements}
	
R.J.T. and S.L. were supported through Australian Research Council grant LP190100621.
	
\end{acknowledgements}

\bibliographystyle{apsrev4-2}

\begin{thebibliography}{52}%
\makeatletter
\providecommand \@ifxundefined [1]{%
 \@ifx{#1\undefined}
}%
\providecommand \@ifnum [1]{%
 \ifnum #1\expandafter \@firstoftwo
 \else \expandafter \@secondoftwo
 \fi
}%
\providecommand \@ifx [1]{%
 \ifx #1\expandafter \@firstoftwo
 \else \expandafter \@secondoftwo
 \fi
}%
\providecommand \natexlab [1]{#1}%
\providecommand \enquote  [1]{``#1''}%
\providecommand \bibnamefont  [1]{#1}%
\providecommand \bibfnamefont [1]{#1}%
\providecommand \citenamefont [1]{#1}%
\providecommand \href@noop [0]{\@secondoftwo}%
\providecommand \href [0]{\begingroup \@sanitize@url \@href}%
\providecommand \@href[1]{\@@startlink{#1}\@@href}%
\providecommand \@@href[1]{\endgroup#1\@@endlink}%
\providecommand \@sanitize@url [0]{\catcode `\\12\catcode `\$12\catcode
  `\&12\catcode `\#12\catcode `\^12\catcode `\_12\catcode `\%12\relax}%
\providecommand \@@startlink[1]{}%
\providecommand \@@endlink[0]{}%
\providecommand \url  [0]{\begingroup\@sanitize@url \@url }%
\providecommand \@url [1]{\endgroup\@href {#1}{\urlprefix }}%
\providecommand \urlprefix  [0]{URL }%
\providecommand \Eprint [0]{\href }%
\providecommand \doibase [0]{https://doi.org/}%
\providecommand \selectlanguage [0]{\@gobble}%
\providecommand \bibinfo  [0]{\@secondoftwo}%
\providecommand \bibfield  [0]{\@secondoftwo}%
\providecommand \translation [1]{[#1]}%
\providecommand \BibitemOpen [0]{}%
\providecommand \bibitemStop [0]{}%
\providecommand \bibitemNoStop [0]{.\EOS\space}%
\providecommand \EOS [0]{\spacefactor3000\relax}%
\providecommand \BibitemShut  [1]{\csname bibitem#1\endcsname}%
\let\auto@bib@innerbib\@empty
\bibitem [{\citenamefont {Bell}(1998)}]{bell_gravity_1998}%
  \BibitemOpen
  \bibfield  {author} {\bibinfo {author} {\bibfnamefont {R.~E.}\ \bibnamefont
  {Bell}},\ }\href {https://www.jstor.org/stable/26057859} {\bibfield
  {journal} {\bibinfo  {journal} {Scientific American}\ }\textbf {\bibinfo
  {volume} {278}},\ \bibinfo {pages} {74} (\bibinfo {year} {1998})}\BibitemShut
  {NoStop}%
\bibitem [{\citenamefont {Pawlowski}(1998)}]{pawlowski_gravity_1998}%
  \BibitemOpen
  \bibfield  {author} {\bibinfo {author} {\bibfnamefont {B.}~\bibnamefont
  {Pawlowski}},\ }\href {https://doi.org/10.1190/1.1437820} {\bibfield
  {journal} {\bibinfo  {journal} {The Leading Edge}\ }\textbf {\bibinfo
  {volume} {17}},\ \bibinfo {pages} {51} (\bibinfo {year} {1998})}\BibitemShut
  {NoStop}%
\bibitem [{\citenamefont {Dransfield}(2007)}]{dransfield_pdf_2007}%
  \BibitemOpen
  \bibfield  {author} {\bibinfo {author} {\bibfnamefont {M.}~\bibnamefont
  {Dransfield}},\ }in\ \href
  {https://www.researchgate.net/publication/267954003_Airborne_Gravity_Gradiometry_in_the_Search_for_Mineral_Deposits}
  {\emph {\bibinfo {booktitle} {Proceedings of {Exploration} 07}}}\ (\bibinfo
  {publisher} {Decennial Mineral Expoloration Conferences},\ \bibinfo {address}
  {Toronto, Canada},\ \bibinfo {year} {2007})\ pp.\ \bibinfo {pages}
  {341--354}\BibitemShut {NoStop}%
\bibitem [{\citenamefont {Salem}\ \emph {et~al.}(2013)\citenamefont {Salem},
  \citenamefont {Masterton}, \citenamefont {Campbell}, \citenamefont
  {Fairhead}, \citenamefont {Dickinson},\ and\ \citenamefont
  {Murphy}}]{salem_interpretation_2013}%
  \BibitemOpen
  \bibfield  {author} {\bibinfo {author} {\bibfnamefont {A.}~\bibnamefont
  {Salem}}, \bibinfo {author} {\bibfnamefont {S.}~\bibnamefont {Masterton}},
  \bibinfo {author} {\bibfnamefont {S.}~\bibnamefont {Campbell}}, \bibinfo
  {author} {\bibfnamefont {J.~D.}\ \bibnamefont {Fairhead}}, \bibinfo {author}
  {\bibfnamefont {J.}~\bibnamefont {Dickinson}},\ and\ \bibinfo {author}
  {\bibfnamefont {C.}~\bibnamefont {Murphy}},\ }\href
  {https://doi.org/10.1111/1365-2478.12039} {\bibfield  {journal} {\bibinfo
  {journal} {Geophysical Prospecting}\ }\textbf {\bibinfo {volume} {61}},\
  \bibinfo {pages} {1065} (\bibinfo {year} {2013})}\BibitemShut {NoStop}%
\bibitem [{\citenamefont {Veryaskin}(2021)}]{veryaskin_gravity_2021}%
  \BibitemOpen
  \bibfield  {author} {\bibinfo {author} {\bibfnamefont {A.~V.}\ \bibnamefont
  {Veryaskin}},\ }in\ \href {https://doi.org/10.1088/978-0-7503-3803-5ch1}
  {\emph {\bibinfo {booktitle} {Gravity, {Magnetic} and {Electromagnetic}
  {Gradiometry} ({Second} {Edition}): {Strategic} technologies in the 21st
  century}}}\ (\bibinfo {year} {2021})\BibitemShut {NoStop}%
\bibitem [{\citenamefont {Wan}\ \emph {et~al.}(2019)\citenamefont {Wan},
  \citenamefont {Ran},\ and\ \citenamefont {Jin}}]{wan_sensitivity_2019}%
  \BibitemOpen
  \bibfield  {author} {\bibinfo {author} {\bibfnamefont {X.}~\bibnamefont
  {Wan}}, \bibinfo {author} {\bibfnamefont {J.}~\bibnamefont {Ran}},\ and\
  \bibinfo {author} {\bibfnamefont {S.}~\bibnamefont {Jin}},\ }\href
  {https://doi.org/10.1007/s11001-018-9361-8} {\bibfield  {journal} {\bibinfo
  {journal} {Marine Geophysical Research}\ }\textbf {\bibinfo {volume} {40}},\
  \bibinfo {pages} {87} (\bibinfo {year} {2019})}\BibitemShut {NoStop}%
\bibitem [{\citenamefont {Xu}\ \emph {et~al.}(2024)\citenamefont {Xu},
  \citenamefont {Yu}, \citenamefont {Zeng}, \citenamefont {Wang}, \citenamefont
  {Tian},\ and\ \citenamefont {Sun}}]{xu_predicting_2024}%
  \BibitemOpen
  \bibfield  {author} {\bibinfo {author} {\bibfnamefont {H.}~\bibnamefont
  {Xu}}, \bibinfo {author} {\bibfnamefont {J.}~\bibnamefont {Yu}}, \bibinfo
  {author} {\bibfnamefont {Y.}~\bibnamefont {Zeng}}, \bibinfo {author}
  {\bibfnamefont {Q.}~\bibnamefont {Wang}}, \bibinfo {author} {\bibfnamefont
  {Y.}~\bibnamefont {Tian}},\ and\ \bibinfo {author} {\bibfnamefont
  {Z.}~\bibnamefont {Sun}},\ }\href
  {https://doi.org/10.1016/j.geog.2023.12.006} {\bibfield  {journal} {\bibinfo
  {journal} {Geodesy and Geodynamics}\ }\textbf {\bibinfo {volume} {15}},\
  \bibinfo {pages} {386} (\bibinfo {year} {2024})}\BibitemShut {NoStop}%
\bibitem [{\citenamefont {Edwards}\ \emph {et~al.}(1997)\citenamefont
  {Edwards}, \citenamefont {Maki},\ and\ \citenamefont
  {Peterson}}]{edwards_gravity_1997}%
  \BibitemOpen
  \bibfield  {author} {\bibinfo {author} {\bibfnamefont {A.~J.}\ \bibnamefont
  {Edwards}}, \bibinfo {author} {\bibfnamefont {J.~T.}\ \bibnamefont {Maki}},\
  and\ \bibinfo {author} {\bibfnamefont {D.~G.}\ \bibnamefont {Peterson}},\
  }\href {https://doi.org/10.4133/JEEG2.2.137} {\bibfield  {journal} {\bibinfo
  {journal} {Journal of Environmental and Engineering Geophysics}\ }\textbf
  {\bibinfo {volume} {2}},\ \bibinfo {pages} {137} (\bibinfo {year}
  {1997})}\BibitemShut {NoStop}%
\bibitem [{\citenamefont {Stray}\ \emph {et~al.}(2022)\citenamefont {Stray},
  \citenamefont {Lamb}, \citenamefont {Kaushik}, \citenamefont {Vovrosh},
  \citenamefont {Rodgers}, \citenamefont {Winch}, \citenamefont {Hayati},
  \citenamefont {Boddice}, \citenamefont {Stabrawa}, \citenamefont {Niggebaum},
  \citenamefont {Langlois}, \citenamefont {Lien}, \citenamefont {Lellouch},
  \citenamefont {Roshanmanesh}, \citenamefont {Ridley}, \citenamefont
  {de~Villiers}, \citenamefont {Brown}, \citenamefont {Cross}, \citenamefont
  {Tuckwell}, \citenamefont {Faramarzi}, \citenamefont {Metje}, \citenamefont
  {Bongs},\ and\ \citenamefont {Holynski}}]{stray_quantum_2022}%
  \BibitemOpen
  \bibfield  {author} {\bibinfo {author} {\bibfnamefont {B.}~\bibnamefont
  {Stray}}, \bibinfo {author} {\bibfnamefont {A.}~\bibnamefont {Lamb}},
  \bibinfo {author} {\bibfnamefont {A.}~\bibnamefont {Kaushik}}, \bibinfo
  {author} {\bibfnamefont {J.}~\bibnamefont {Vovrosh}}, \bibinfo {author}
  {\bibfnamefont {A.}~\bibnamefont {Rodgers}}, \bibinfo {author} {\bibfnamefont
  {J.}~\bibnamefont {Winch}}, \bibinfo {author} {\bibfnamefont
  {F.}~\bibnamefont {Hayati}}, \bibinfo {author} {\bibfnamefont
  {D.}~\bibnamefont {Boddice}}, \bibinfo {author} {\bibfnamefont
  {A.}~\bibnamefont {Stabrawa}}, \bibinfo {author} {\bibfnamefont
  {A.}~\bibnamefont {Niggebaum}}, \bibinfo {author} {\bibfnamefont
  {M.}~\bibnamefont {Langlois}}, \bibinfo {author} {\bibfnamefont {Y.-H.}\
  \bibnamefont {Lien}}, \bibinfo {author} {\bibfnamefont {S.}~\bibnamefont
  {Lellouch}}, \bibinfo {author} {\bibfnamefont {S.}~\bibnamefont
  {Roshanmanesh}}, \bibinfo {author} {\bibfnamefont {K.}~\bibnamefont
  {Ridley}}, \bibinfo {author} {\bibfnamefont {G.}~\bibnamefont {de~Villiers}},
  \bibinfo {author} {\bibfnamefont {G.}~\bibnamefont {Brown}}, \bibinfo
  {author} {\bibfnamefont {T.}~\bibnamefont {Cross}}, \bibinfo {author}
  {\bibfnamefont {G.}~\bibnamefont {Tuckwell}}, \bibinfo {author}
  {\bibfnamefont {A.}~\bibnamefont {Faramarzi}}, \bibinfo {author}
  {\bibfnamefont {N.}~\bibnamefont {Metje}}, \bibinfo {author} {\bibfnamefont
  {K.}~\bibnamefont {Bongs}},\ and\ \bibinfo {author} {\bibfnamefont
  {M.}~\bibnamefont {Holynski}},\ }\href
  {https://doi.org/10.1038/s41586-021-04315-3} {\bibfield  {journal} {\bibinfo
  {journal} {Nature}\ }\textbf {\bibinfo {volume} {602}},\ \bibinfo {pages}
  {590} (\bibinfo {year} {2022})}\BibitemShut {NoStop}%
\bibitem [{\citenamefont {Qiao}\ \emph {et~al.}(2025)\citenamefont {Qiao},
  \citenamefont {Zhang}, \citenamefont {Yuan}, \citenamefont {Li},
  \citenamefont {Hu}, \citenamefont {Yang}, \citenamefont {Weng}, \citenamefont
  {Zhang}, \citenamefont {Wang}, \citenamefont {Qin}, \citenamefont {Hao},
  \citenamefont {Yang}, \citenamefont {Lv}, \citenamefont {Wu}, \citenamefont
  {Wu}, \citenamefont {Wang},\ and\ \citenamefont
  {Lin}}]{qiao_application_2025}%
  \BibitemOpen
  \bibfield  {author} {\bibinfo {author} {\bibfnamefont {Z.-k.}\ \bibnamefont
  {Qiao}}, \bibinfo {author} {\bibfnamefont {J.-j.}\ \bibnamefont {Zhang}},
  \bibinfo {author} {\bibfnamefont {P.}~\bibnamefont {Yuan}}, \bibinfo {author}
  {\bibfnamefont {L.-l.}\ \bibnamefont {Li}}, \bibinfo {author} {\bibfnamefont
  {R.}~\bibnamefont {Hu}}, \bibinfo {author} {\bibfnamefont {H.-x.}\
  \bibnamefont {Yang}}, \bibinfo {author} {\bibfnamefont {K.-x.}\ \bibnamefont
  {Weng}}, \bibinfo {author} {\bibfnamefont {Z.-y.}\ \bibnamefont {Zhang}},
  \bibinfo {author} {\bibfnamefont {L.-f.}\ \bibnamefont {Wang}}, \bibinfo
  {author} {\bibfnamefont {P.}~\bibnamefont {Qin}}, \bibinfo {author}
  {\bibfnamefont {L.-l.}\ \bibnamefont {Hao}}, \bibinfo {author} {\bibfnamefont
  {C.}~\bibnamefont {Yang}}, \bibinfo {author} {\bibfnamefont {X.-f.}\
  \bibnamefont {Lv}}, \bibinfo {author} {\bibfnamefont {X.-m.}\ \bibnamefont
  {Wu}}, \bibinfo {author} {\bibfnamefont {B.}~\bibnamefont {Wu}}, \bibinfo
  {author} {\bibfnamefont {X.-l.}\ \bibnamefont {Wang}},\ and\ \bibinfo
  {author} {\bibfnamefont {Q.}~\bibnamefont {Lin}},\ }\href
  {https://doi.org/10.1016/j.jappgeo.2025.105700} {\bibfield  {journal}
  {\bibinfo  {journal} {Journal of Applied Geophysics}\ }\textbf {\bibinfo
  {volume} {237}},\ \bibinfo {pages} {105700} (\bibinfo {year}
  {2025})}\BibitemShut {NoStop}%
\bibitem [{\citenamefont {Rummel}\ \emph {et~al.}(2011)\citenamefont {Rummel},
  \citenamefont {Yi},\ and\ \citenamefont {Stummer}}]{rummel_goce_2011}%
  \BibitemOpen
  \bibfield  {author} {\bibinfo {author} {\bibfnamefont {R.}~\bibnamefont
  {Rummel}}, \bibinfo {author} {\bibfnamefont {W.}~\bibnamefont {Yi}},\ and\
  \bibinfo {author} {\bibfnamefont {C.}~\bibnamefont {Stummer}},\ }\href
  {https://doi.org/10.1007/s00190-011-0500-0} {\bibfield  {journal} {\bibinfo
  {journal} {Journal of Geodesy}\ }\textbf {\bibinfo {volume} {85}},\ \bibinfo
  {pages} {777} (\bibinfo {year} {2011})}\BibitemShut {NoStop}%
\bibitem [{\citenamefont {Kebede}\ and\ \citenamefont
  {Mammo}(2021)}]{kebede_processing_2021}%
  \BibitemOpen
  \bibfield  {author} {\bibinfo {author} {\bibfnamefont {B.}~\bibnamefont
  {Kebede}}\ and\ \bibinfo {author} {\bibfnamefont {T.}~\bibnamefont {Mammo}},\
  }\href {https://doi.org/10.1016/j.heliyon.2021.e06872} {\bibfield  {journal}
  {\bibinfo  {journal} {Heliyon}\ }\textbf {\bibinfo {volume} {7}},\ \bibinfo
  {pages} {e06872} (\bibinfo {year} {2021})}\BibitemShut {NoStop}%
\bibitem [{\citenamefont {Snadden}\ \emph {et~al.}(1998)\citenamefont
  {Snadden}, \citenamefont {McGuirk}, \citenamefont {Bouyer}, \citenamefont
  {Haritos},\ and\ \citenamefont {Kasevich}}]{snadden_measurement_1998}%
  \BibitemOpen
  \bibfield  {author} {\bibinfo {author} {\bibfnamefont {M.~J.}\ \bibnamefont
  {Snadden}}, \bibinfo {author} {\bibfnamefont {J.~M.}\ \bibnamefont
  {McGuirk}}, \bibinfo {author} {\bibfnamefont {P.}~\bibnamefont {Bouyer}},
  \bibinfo {author} {\bibfnamefont {K.~G.}\ \bibnamefont {Haritos}},\ and\
  \bibinfo {author} {\bibfnamefont {M.~A.}\ \bibnamefont {Kasevich}},\ }\href
  {https://doi.org/10.1103/PhysRevLett.81.971} {\bibfield  {journal} {\bibinfo
  {journal} {Physical Review Letters}\ }\textbf {\bibinfo {volume} {81}},\
  \bibinfo {pages} {971} (\bibinfo {year} {1998})}\BibitemShut {NoStop}%
\bibitem [{\citenamefont {Fixler}\ \emph {et~al.}(2007)\citenamefont {Fixler},
  \citenamefont {Foster}, \citenamefont {McGuirk},\ and\ \citenamefont
  {Kasevich}}]{fixler_atom_2007}%
  \BibitemOpen
  \bibfield  {author} {\bibinfo {author} {\bibfnamefont {J.~B.}\ \bibnamefont
  {Fixler}}, \bibinfo {author} {\bibfnamefont {G.~T.}\ \bibnamefont {Foster}},
  \bibinfo {author} {\bibfnamefont {J.~M.}\ \bibnamefont {McGuirk}},\ and\
  \bibinfo {author} {\bibfnamefont {M.~A.}\ \bibnamefont {Kasevich}},\ }\href
  {https://doi.org/10.1126/science.1135459} {\bibfield  {journal} {\bibinfo
  {journal} {Science}\ }\textbf {\bibinfo {volume} {315}},\ \bibinfo {pages}
  {74} (\bibinfo {year} {2007})}\BibitemShut {NoStop}%
\bibitem [{\citenamefont {Rosi}\ \emph {et~al.}(2014)\citenamefont {Rosi},
  \citenamefont {Sorrentino}, \citenamefont {Cacciapuoti}, \citenamefont
  {Prevedelli},\ and\ \citenamefont {Tino}}]{rosi_precision_2014}%
  \BibitemOpen
  \bibfield  {author} {\bibinfo {author} {\bibfnamefont {G.}~\bibnamefont
  {Rosi}}, \bibinfo {author} {\bibfnamefont {F.}~\bibnamefont {Sorrentino}},
  \bibinfo {author} {\bibfnamefont {L.}~\bibnamefont {Cacciapuoti}}, \bibinfo
  {author} {\bibfnamefont {M.}~\bibnamefont {Prevedelli}},\ and\ \bibinfo
  {author} {\bibfnamefont {G.~M.}\ \bibnamefont {Tino}},\ }\href
  {https://doi.org/10.1038/nature13433} {\bibfield  {journal} {\bibinfo
  {journal} {Nature}\ }\textbf {\bibinfo {volume} {510}},\ \bibinfo {pages}
  {518} (\bibinfo {year} {2014})}\BibitemShut {NoStop}%
\bibitem [{\citenamefont {Asenbaum}\ \emph {et~al.}(2017)\citenamefont
  {Asenbaum}, \citenamefont {Overstreet}, \citenamefont {Kovachy},
  \citenamefont {Brown}, \citenamefont {Hogan},\ and\ \citenamefont
  {Kasevich}}]{asenbaum_phase_2017}%
  \BibitemOpen
  \bibfield  {author} {\bibinfo {author} {\bibfnamefont {P.}~\bibnamefont
  {Asenbaum}}, \bibinfo {author} {\bibfnamefont {C.}~\bibnamefont
  {Overstreet}}, \bibinfo {author} {\bibfnamefont {T.}~\bibnamefont {Kovachy}},
  \bibinfo {author} {\bibfnamefont {D.~D.}\ \bibnamefont {Brown}}, \bibinfo
  {author} {\bibfnamefont {J.~M.}\ \bibnamefont {Hogan}},\ and\ \bibinfo
  {author} {\bibfnamefont {M.~A.}\ \bibnamefont {Kasevich}},\ }\href
  {https://doi.org/10.1103/PhysRevLett.118.183602} {\bibfield  {journal}
  {\bibinfo  {journal} {Physical Review Letters}\ }\textbf {\bibinfo {volume}
  {118}},\ \bibinfo {pages} {183602} (\bibinfo {year} {2017})}\BibitemShut
  {NoStop}%
\bibitem [{\citenamefont {Asenbaum}\ \emph {et~al.}(2020)\citenamefont
  {Asenbaum}, \citenamefont {Overstreet}, \citenamefont {Kim}, \citenamefont
  {Curti},\ and\ \citenamefont
  {Kasevich}}]{asenbaum_atom-interferometric_2020}%
  \BibitemOpen
  \bibfield  {author} {\bibinfo {author} {\bibfnamefont {P.}~\bibnamefont
  {Asenbaum}}, \bibinfo {author} {\bibfnamefont {C.}~\bibnamefont
  {Overstreet}}, \bibinfo {author} {\bibfnamefont {M.}~\bibnamefont {Kim}},
  \bibinfo {author} {\bibfnamefont {J.}~\bibnamefont {Curti}},\ and\ \bibinfo
  {author} {\bibfnamefont {M.~A.}\ \bibnamefont {Kasevich}},\ }\href
  {https://doi.org/10.1103/PhysRevLett.125.191101} {\bibfield  {journal}
  {\bibinfo  {journal} {Physical Review Letters}\ }\textbf {\bibinfo {volume}
  {125}},\ \bibinfo {pages} {191101} (\bibinfo {year} {2020})}\BibitemShut
  {NoStop}%
\bibitem [{\citenamefont {Barrett}\ \emph {et~al.}(2015)\citenamefont
  {Barrett}, \citenamefont {Antoni-Micollier}, \citenamefont {Chichet},
  \citenamefont {Battelier}, \citenamefont {Gominet}, \citenamefont {Bertoldi},
  \citenamefont {Bouyer},\ and\ \citenamefont
  {Landragin}}]{barrett_correlative_2015}%
  \BibitemOpen
  \bibfield  {author} {\bibinfo {author} {\bibfnamefont {B.}~\bibnamefont
  {Barrett}}, \bibinfo {author} {\bibfnamefont {L.}~\bibnamefont
  {Antoni-Micollier}}, \bibinfo {author} {\bibfnamefont {L.}~\bibnamefont
  {Chichet}}, \bibinfo {author} {\bibfnamefont {B.}~\bibnamefont {Battelier}},
  \bibinfo {author} {\bibfnamefont {P.-A.}\ \bibnamefont {Gominet}}, \bibinfo
  {author} {\bibfnamefont {A.}~\bibnamefont {Bertoldi}}, \bibinfo {author}
  {\bibfnamefont {P.}~\bibnamefont {Bouyer}},\ and\ \bibinfo {author}
  {\bibfnamefont {A.}~\bibnamefont {Landragin}},\ }\href
  {https://doi.org/10.1088/1367-2630/17/8/085010} {\bibfield  {journal}
  {\bibinfo  {journal} {New Journal of Physics}\ }\textbf {\bibinfo {volume}
  {17}},\ \bibinfo {pages} {085010} (\bibinfo {year} {2015})}\BibitemShut
  {NoStop}%
\bibitem [{\citenamefont {Weiner}\ \emph {et~al.}(2020)\citenamefont {Weiner},
  \citenamefont {Wu}, \citenamefont {Pagel}, \citenamefont {Li}, \citenamefont
  {Sleczkowski}, \citenamefont {Ketcham},\ and\ \citenamefont
  {Mueller}}]{weiner_flight_2020}%
  \BibitemOpen
  \bibfield  {author} {\bibinfo {author} {\bibfnamefont {S.}~\bibnamefont
  {Weiner}}, \bibinfo {author} {\bibfnamefont {X.}~\bibnamefont {Wu}}, \bibinfo
  {author} {\bibfnamefont {Z.}~\bibnamefont {Pagel}}, \bibinfo {author}
  {\bibfnamefont {D.}~\bibnamefont {Li}}, \bibinfo {author} {\bibfnamefont
  {J.}~\bibnamefont {Sleczkowski}}, \bibinfo {author} {\bibfnamefont
  {F.}~\bibnamefont {Ketcham}},\ and\ \bibinfo {author} {\bibfnamefont
  {H.}~\bibnamefont {Mueller}},\ }in\ \href
  {https://doi.org/10.1109/INERTIAL48129.2020.9090014} {\emph {\bibinfo
  {booktitle} {2020 {IEEE} {International} {Symposium} on {Inertial} {Sensors}
  and {Systems} ({INERTIAL})}}}\ (\bibinfo {address} {Hiroshima, Japan},\
  \bibinfo {year} {2020})\ pp.\ \bibinfo {pages} {1--3}\BibitemShut {NoStop}%
\bibitem [{\citenamefont {Janvier}\ \emph {et~al.}(2022)\citenamefont
  {Janvier}, \citenamefont {Ménoret}, \citenamefont {Desruelle}, \citenamefont
  {Merlet}, \citenamefont {Landragin},\ and\ \citenamefont {Pereira~dos
  Santos}}]{janvier_compact_2022}%
  \BibitemOpen
  \bibfield  {author} {\bibinfo {author} {\bibfnamefont {C.}~\bibnamefont
  {Janvier}}, \bibinfo {author} {\bibfnamefont {V.}~\bibnamefont {Ménoret}},
  \bibinfo {author} {\bibfnamefont {B.}~\bibnamefont {Desruelle}}, \bibinfo
  {author} {\bibfnamefont {S.}~\bibnamefont {Merlet}}, \bibinfo {author}
  {\bibfnamefont {A.}~\bibnamefont {Landragin}},\ and\ \bibinfo {author}
  {\bibfnamefont {F.}~\bibnamefont {Pereira~dos Santos}},\ }\href
  {https://doi.org/10.1103/PhysRevA.105.022801} {\bibfield  {journal} {\bibinfo
   {journal} {Physical Review A}\ }\textbf {\bibinfo {volume} {105}},\ \bibinfo
  {pages} {022801} (\bibinfo {year} {2022})}\BibitemShut {NoStop}%
\bibitem [{\citenamefont {Lyu}\ \emph {et~al.}(2022)\citenamefont {Lyu},
  \citenamefont {Zhong}, \citenamefont {Zhang}, \citenamefont {Liu},
  \citenamefont {Zhu}, \citenamefont {Xu}, \citenamefont {Chen}, \citenamefont
  {Tang}, \citenamefont {Wang},\ and\ \citenamefont {Zhan}}]{lyu_compact_2022}%
  \BibitemOpen
  \bibfield  {author} {\bibinfo {author} {\bibfnamefont {W.}~\bibnamefont
  {Lyu}}, \bibinfo {author} {\bibfnamefont {J.-Q.}\ \bibnamefont {Zhong}},
  \bibinfo {author} {\bibfnamefont {X.-W.}\ \bibnamefont {Zhang}}, \bibinfo
  {author} {\bibfnamefont {W.}~\bibnamefont {Liu}}, \bibinfo {author}
  {\bibfnamefont {L.}~\bibnamefont {Zhu}}, \bibinfo {author} {\bibfnamefont
  {W.-H.}\ \bibnamefont {Xu}}, \bibinfo {author} {\bibfnamefont
  {X.}~\bibnamefont {Chen}}, \bibinfo {author} {\bibfnamefont {B.}~\bibnamefont
  {Tang}}, \bibinfo {author} {\bibfnamefont {J.}~\bibnamefont {Wang}},\ and\
  \bibinfo {author} {\bibfnamefont {M.-S.}\ \bibnamefont {Zhan}},\ }\href
  {https://doi.org/10.1103/PhysRevApplied.18.054091} {\bibfield  {journal}
  {\bibinfo  {journal} {Physical Review Applied}\ }\textbf {\bibinfo {volume}
  {18}},\ \bibinfo {pages} {054091} (\bibinfo {year} {2022})}\BibitemShut
  {NoStop}%
\bibitem [{del()}]{delta_g_company_2025}%
  \BibitemOpen
  \href
  {https://www.birmingham.ac.uk/news/2023/quantum-sensor-for-gravity-gradiometry-successfully-validated-at-sea}
  {\bibinfo {title} {Quantum sensor for gravity successfully validated at
  sea}},\ \bibinfo {note}
  {https://www.birmingham.ac.uk/news/2023/quantum-sensor-for-gravity-gradiometry-successfully-validated-at-sea}\BibitemShut
  {NoStop}%
\bibitem [{\citenamefont {Lellouch}\ and\ \citenamefont
  {Holynski}(2025)}]{lellouch_integration_2025}%
  \BibitemOpen
  \bibfield  {author} {\bibinfo {author} {\bibfnamefont {S.}~\bibnamefont
  {Lellouch}}\ and\ \bibinfo {author} {\bibfnamefont {M.}~\bibnamefont
  {Holynski}},\ }\href {https://doi.org/10.1088/2058-9565/adf2d9} {\bibfield
  {journal} {\bibinfo  {journal} {Quantum Science and Technology}\ }\textbf
  {\bibinfo {volume} {10}},\ \bibinfo {pages} {045007} (\bibinfo {year}
  {2025})}\BibitemShut {NoStop}%
\bibitem [{\citenamefont {Biedermann}\ \emph {et~al.}(2015)\citenamefont
  {Biedermann}, \citenamefont {Wu}, \citenamefont {Deslauriers}, \citenamefont
  {Roy}, \citenamefont {Mahadeswaraswamy},\ and\ \citenamefont
  {Kasevich}}]{biedermann_testing_2015}%
  \BibitemOpen
  \bibfield  {author} {\bibinfo {author} {\bibfnamefont {G.~W.}\ \bibnamefont
  {Biedermann}}, \bibinfo {author} {\bibfnamefont {X.}~\bibnamefont {Wu}},
  \bibinfo {author} {\bibfnamefont {L.}~\bibnamefont {Deslauriers}}, \bibinfo
  {author} {\bibfnamefont {S.}~\bibnamefont {Roy}}, \bibinfo {author}
  {\bibfnamefont {C.}~\bibnamefont {Mahadeswaraswamy}},\ and\ \bibinfo {author}
  {\bibfnamefont {M.~A.}\ \bibnamefont {Kasevich}},\ }\href
  {https://doi.org/10.1103/PhysRevA.91.033629} {\bibfield  {journal} {\bibinfo
  {journal} {Physical Review A}\ }\textbf {\bibinfo {volume} {91}},\ \bibinfo
  {pages} {033629} (\bibinfo {year} {2015})}\BibitemShut {NoStop}%
\bibitem [{\citenamefont {Stolzenberg}\ \emph {et~al.}(2025)\citenamefont
  {Stolzenberg}, \citenamefont {Struckmann}, \citenamefont {Bode},
  \citenamefont {Li}, \citenamefont {Herbst}, \citenamefont {Vollenkemper},
  \citenamefont {Thomas}, \citenamefont {Rajagopalan}, \citenamefont {Rasel},
  \citenamefont {Gaaloul},\ and\ \citenamefont
  {Schlippert}}]{stolzenberg_multi-axis_2025}%
  \BibitemOpen
  \bibfield  {author} {\bibinfo {author} {\bibfnamefont {K.}~\bibnamefont
  {Stolzenberg}}, \bibinfo {author} {\bibfnamefont {C.}~\bibnamefont
  {Struckmann}}, \bibinfo {author} {\bibfnamefont {S.}~\bibnamefont {Bode}},
  \bibinfo {author} {\bibfnamefont {R.}~\bibnamefont {Li}}, \bibinfo {author}
  {\bibfnamefont {A.}~\bibnamefont {Herbst}}, \bibinfo {author} {\bibfnamefont
  {V.}~\bibnamefont {Vollenkemper}}, \bibinfo {author} {\bibfnamefont
  {D.}~\bibnamefont {Thomas}}, \bibinfo {author} {\bibfnamefont
  {A.}~\bibnamefont {Rajagopalan}}, \bibinfo {author} {\bibfnamefont
  {E.}~\bibnamefont {Rasel}}, \bibinfo {author} {\bibfnamefont
  {N.}~\bibnamefont {Gaaloul}},\ and\ \bibinfo {author} {\bibfnamefont
  {D.}~\bibnamefont {Schlippert}},\ }\href
  {https://doi.org/10.1103/PhysRevLett.134.143601} {\bibfield  {journal}
  {\bibinfo  {journal} {Physical Review Letters}\ }\textbf {\bibinfo {volume}
  {134}},\ \bibinfo {pages} {143601} (\bibinfo {year} {2025})}\BibitemShut
  {NoStop}%
\bibitem [{\citenamefont {Carraz}\ \emph {et~al.}(2014)\citenamefont {Carraz},
  \citenamefont {Siemes}, \citenamefont {Massotti}, \citenamefont {Haagmans},\
  and\ \citenamefont {Silvestrin}}]{carraz_spaceborne_2014}%
  \BibitemOpen
  \bibfield  {author} {\bibinfo {author} {\bibfnamefont {O.}~\bibnamefont
  {Carraz}}, \bibinfo {author} {\bibfnamefont {C.}~\bibnamefont {Siemes}},
  \bibinfo {author} {\bibfnamefont {L.}~\bibnamefont {Massotti}}, \bibinfo
  {author} {\bibfnamefont {R.}~\bibnamefont {Haagmans}},\ and\ \bibinfo
  {author} {\bibfnamefont {P.}~\bibnamefont {Silvestrin}},\ }\href
  {https://doi.org/10.1007/s12217-014-9385-x} {\bibfield  {journal} {\bibinfo
  {journal} {Microgravity Science and Technology}\ }\textbf {\bibinfo {volume}
  {26}},\ \bibinfo {pages} {139} (\bibinfo {year} {2014})}\BibitemShut
  {NoStop}%
\bibitem [{\citenamefont {Duan}\ \emph {et~al.}(2014)\citenamefont {Duan},
  \citenamefont {Zhou}, \citenamefont {Mao}, \citenamefont {Yao}, \citenamefont
  {Deng}, \citenamefont {Luo},\ and\ \citenamefont {Hu}}]{duan_operating_2014}%
  \BibitemOpen
  \bibfield  {author} {\bibinfo {author} {\bibfnamefont {X.-C.}\ \bibnamefont
  {Duan}}, \bibinfo {author} {\bibfnamefont {M.-K.}\ \bibnamefont {Zhou}},
  \bibinfo {author} {\bibfnamefont {D.-K.}\ \bibnamefont {Mao}}, \bibinfo
  {author} {\bibfnamefont {H.-B.}\ \bibnamefont {Yao}}, \bibinfo {author}
  {\bibfnamefont {X.-B.}\ \bibnamefont {Deng}}, \bibinfo {author}
  {\bibfnamefont {J.}~\bibnamefont {Luo}},\ and\ \bibinfo {author}
  {\bibfnamefont {Z.-K.}\ \bibnamefont {Hu}},\ }\href
  {https://doi.org/10.1103/PhysRevA.90.023617} {\bibfield  {journal} {\bibinfo
  {journal} {Physical Review A}\ }\textbf {\bibinfo {volume} {90}},\ \bibinfo
  {pages} {023617} (\bibinfo {year} {2014})}\BibitemShut {NoStop}%
\bibitem [{Note1()}]{Note1}%
  \BibitemOpen
  \bibinfo {note} {Since the gravitational field $\protect \mathbf {g}$ is the
  gradient of a potential $V(\protect \mathbf {r})$, symmetry in $\protect
  \mathrm {G}$ is due to the commutativity of partial derivatives. $\protect
  \mathrm {G}$ is traceless in free space since $\nabla ^2 V(\protect \mathbf
  {r}) = 0$.}\BibitemShut {Stop}%
\bibitem [{\citenamefont {Trimeche}\ \emph {et~al.}(2017)\citenamefont
  {Trimeche}, \citenamefont {Langlois}, \citenamefont {Merlet},\ and\
  \citenamefont {Pereira Dos~Santos}}]{trimeche_active_2017}%
  \BibitemOpen
  \bibfield  {author} {\bibinfo {author} {\bibfnamefont {A.}~\bibnamefont
  {Trimeche}}, \bibinfo {author} {\bibfnamefont {M.}~\bibnamefont {Langlois}},
  \bibinfo {author} {\bibfnamefont {S.}~\bibnamefont {Merlet}},\ and\ \bibinfo
  {author} {\bibfnamefont {F.}~\bibnamefont {Pereira Dos~Santos}},\ }\href
  {https://doi.org/10.1103/PhysRevApplied.7.034016} {\bibfield  {journal}
  {\bibinfo  {journal} {Physical Review Applied}\ }\textbf {\bibinfo {volume}
  {7}},\ \bibinfo {pages} {034016} (\bibinfo {year} {2017})}\BibitemShut
  {NoStop}%
\bibitem [{\citenamefont {Schkolnik}\ \emph {et~al.}(2015)\citenamefont
  {Schkolnik}, \citenamefont {Leykauf}, \citenamefont {Hauth}, \citenamefont
  {Freier},\ and\ \citenamefont {Peters}}]{schkolnik_effect_2015}%
  \BibitemOpen
  \bibfield  {author} {\bibinfo {author} {\bibfnamefont {V.}~\bibnamefont
  {Schkolnik}}, \bibinfo {author} {\bibfnamefont {B.}~\bibnamefont {Leykauf}},
  \bibinfo {author} {\bibfnamefont {M.}~\bibnamefont {Hauth}}, \bibinfo
  {author} {\bibfnamefont {C.}~\bibnamefont {Freier}},\ and\ \bibinfo {author}
  {\bibfnamefont {A.}~\bibnamefont {Peters}},\ }\href
  {https://doi.org/10.1007/s00340-015-6138-5} {\bibfield  {journal} {\bibinfo
  {journal} {Applied Physics B}\ }\textbf {\bibinfo {volume} {120}},\ \bibinfo
  {pages} {311} (\bibinfo {year} {2015})}\BibitemShut {NoStop}%
\bibitem [{\citenamefont {Roura}\ \emph {et~al.}(2014)\citenamefont {Roura},
  \citenamefont {Zeller},\ and\ \citenamefont
  {Schleich}}]{roura_overcoming_2014}%
  \BibitemOpen
  \bibfield  {author} {\bibinfo {author} {\bibfnamefont {A.}~\bibnamefont
  {Roura}}, \bibinfo {author} {\bibfnamefont {W.}~\bibnamefont {Zeller}},\ and\
  \bibinfo {author} {\bibfnamefont {W.~P.}\ \bibnamefont {Schleich}},\ }\href
  {https://doi.org/10.1088/1367-2630/16/12/123012} {\bibfield  {journal}
  {\bibinfo  {journal} {New Journal of Physics}\ }\textbf {\bibinfo {volume}
  {16}},\ \bibinfo {pages} {123012} (\bibinfo {year} {2014})}\BibitemShut
  {NoStop}%
\bibitem [{\citenamefont {Thomas}(2025)}]{github}%
  \BibitemOpen
  \bibfield  {author} {\bibinfo {author} {\bibfnamefont {R.~J.}\ \bibnamefont
  {Thomas}},\ }\href
  {https://github.com/ryan-james-thomas/interferometer-simulation-bare}
  {\bibinfo {title} {Software for simulating an atom interferometer}},\
  \bibinfo {howpublished}
  {https://github.com/ryan-james-thomas/interferometer-simulation-bare}
  (\bibinfo {year} {2025})\BibitemShut {NoStop}%
\bibitem [{\citenamefont {Gauguet}\ \emph {et~al.}(2008)\citenamefont
  {Gauguet}, \citenamefont {Mehlstäubler}, \citenamefont {Lévèque},
  \citenamefont {Le~Gouët}, \citenamefont {Chaibi}, \citenamefont {Canuel},
  \citenamefont {Clairon}, \citenamefont {Dos~Santos},\ and\ \citenamefont
  {Landragin}}]{gauguet_off-resonant_2008}%
  \BibitemOpen
  \bibfield  {author} {\bibinfo {author} {\bibfnamefont {A.}~\bibnamefont
  {Gauguet}}, \bibinfo {author} {\bibfnamefont {T.~E.}\ \bibnamefont
  {Mehlstäubler}}, \bibinfo {author} {\bibfnamefont {T.}~\bibnamefont
  {Lévèque}}, \bibinfo {author} {\bibfnamefont {J.}~\bibnamefont
  {Le~Gouët}}, \bibinfo {author} {\bibfnamefont {W.}~\bibnamefont {Chaibi}},
  \bibinfo {author} {\bibfnamefont {B.}~\bibnamefont {Canuel}}, \bibinfo
  {author} {\bibfnamefont {A.}~\bibnamefont {Clairon}}, \bibinfo {author}
  {\bibfnamefont {F.~P.}\ \bibnamefont {Dos~Santos}},\ and\ \bibinfo {author}
  {\bibfnamefont {A.}~\bibnamefont {Landragin}},\ }\href
  {https://doi.org/10.1103/PhysRevA.78.043615} {\bibfield  {journal} {\bibinfo
  {journal} {Physical Review A}\ }\textbf {\bibinfo {volume} {78}},\ \bibinfo
  {pages} {043615} (\bibinfo {year} {2008})}\BibitemShut {NoStop}%
\bibitem [{\citenamefont {Hu}\ \emph {et~al.}(2017)\citenamefont {Hu},
  \citenamefont {Chen}, \citenamefont {Fang}, \citenamefont {Zhou},
  \citenamefont {Zhong}, \citenamefont {Wang},\ and\ \citenamefont
  {Zhan}}]{hu_analysis_2017}%
  \BibitemOpen
  \bibfield  {author} {\bibinfo {author} {\bibfnamefont {J.}~\bibnamefont
  {Hu}}, \bibinfo {author} {\bibfnamefont {X.}~\bibnamefont {Chen}}, \bibinfo
  {author} {\bibfnamefont {J.}~\bibnamefont {Fang}}, \bibinfo {author}
  {\bibfnamefont {L.}~\bibnamefont {Zhou}}, \bibinfo {author} {\bibfnamefont
  {J.}~\bibnamefont {Zhong}}, \bibinfo {author} {\bibfnamefont
  {J.}~\bibnamefont {Wang}},\ and\ \bibinfo {author} {\bibfnamefont
  {M.}~\bibnamefont {Zhan}},\ }\href
  {https://doi.org/10.1103/PhysRevA.96.023618} {\bibfield  {journal} {\bibinfo
  {journal} {Physical Review A}\ }\textbf {\bibinfo {volume} {96}},\ \bibinfo
  {pages} {023618} (\bibinfo {year} {2017})}\BibitemShut {NoStop}%
\bibitem [{\citenamefont {d’Armagnac~de Castanet}\ \emph
  {et~al.}(2024)\citenamefont {d’Armagnac~de Castanet}, \citenamefont
  {Des~Cognets}, \citenamefont {Arguel}, \citenamefont {Templier},
  \citenamefont {Jarlaud}, \citenamefont {Ménoret}, \citenamefont {Desruelle},
  \citenamefont {Bouyer},\ and\ \citenamefont
  {Battelier}}]{darmagnac_de_castanet_atom_2024}%
  \BibitemOpen
  \bibfield  {author} {\bibinfo {author} {\bibfnamefont {Q.}~\bibnamefont
  {d’Armagnac~de Castanet}}, \bibinfo {author} {\bibfnamefont
  {C.}~\bibnamefont {Des~Cognets}}, \bibinfo {author} {\bibfnamefont
  {R.}~\bibnamefont {Arguel}}, \bibinfo {author} {\bibfnamefont
  {S.}~\bibnamefont {Templier}}, \bibinfo {author} {\bibfnamefont
  {V.}~\bibnamefont {Jarlaud}}, \bibinfo {author} {\bibfnamefont
  {V.}~\bibnamefont {Ménoret}}, \bibinfo {author} {\bibfnamefont
  {B.}~\bibnamefont {Desruelle}}, \bibinfo {author} {\bibfnamefont
  {P.}~\bibnamefont {Bouyer}},\ and\ \bibinfo {author} {\bibfnamefont
  {B.}~\bibnamefont {Battelier}},\ }\href
  {https://doi.org/10.1038/s41467-024-50804-0} {\bibfield  {journal} {\bibinfo
  {journal} {Nature Communications}\ }\textbf {\bibinfo {volume} {15}},\
  \bibinfo {pages} {6406} (\bibinfo {year} {2024})}\BibitemShut {NoStop}%
\bibitem [{Note2()}]{Note2}%
  \BibitemOpen
  \bibinfo {note} {The contribution of the Earth's background gradient of
  $\mathord {\approx }\protect \,\SI {3000}{\eotvos }$ can be subtracted from
  the measurement, leaving only signals from gravitational
  anomalies}\BibitemShut {NoStop}%
\bibitem [{\citenamefont {Chan}\ and\ \citenamefont
  {Paik}(1987)}]{chan_superconducting_1987}%
  \BibitemOpen
  \bibfield  {author} {\bibinfo {author} {\bibfnamefont {H.~A.}\ \bibnamefont
  {Chan}}\ and\ \bibinfo {author} {\bibfnamefont {H.~J.}\ \bibnamefont
  {Paik}},\ }\href {https://doi.org/10.1103/PhysRevD.35.3551} {\bibfield
  {journal} {\bibinfo  {journal} {Physical Review D}\ }\textbf {\bibinfo
  {volume} {35}},\ \bibinfo {pages} {3551} (\bibinfo {year}
  {1987})}\BibitemShut {NoStop}%
\bibitem [{\citenamefont {Marzlin}\ and\ \citenamefont
  {Audretsch}(1996)}]{marzlin_state_1996}%
  \BibitemOpen
  \bibfield  {author} {\bibinfo {author} {\bibfnamefont {K.-P.}\ \bibnamefont
  {Marzlin}}\ and\ \bibinfo {author} {\bibfnamefont {J.}~\bibnamefont
  {Audretsch}},\ }\href {https://doi.org/10.1103/PhysRevA.53.312} {\bibfield
  {journal} {\bibinfo  {journal} {Physical Review A}\ }\textbf {\bibinfo
  {volume} {53}},\ \bibinfo {pages} {312} (\bibinfo {year} {1996})}\BibitemShut
  {NoStop}%
\bibitem [{\citenamefont {McGuirk}\ \emph {et~al.}(2002)\citenamefont
  {McGuirk}, \citenamefont {Foster}, \citenamefont {Fixler}, \citenamefont
  {Snadden},\ and\ \citenamefont {Kasevich}}]{mcguirk_sensitive_2002}%
  \BibitemOpen
  \bibfield  {author} {\bibinfo {author} {\bibfnamefont {J.~M.}\ \bibnamefont
  {McGuirk}}, \bibinfo {author} {\bibfnamefont {G.~T.}\ \bibnamefont {Foster}},
  \bibinfo {author} {\bibfnamefont {J.~B.}\ \bibnamefont {Fixler}}, \bibinfo
  {author} {\bibfnamefont {M.~J.}\ \bibnamefont {Snadden}},\ and\ \bibinfo
  {author} {\bibfnamefont {M.~A.}\ \bibnamefont {Kasevich}},\ }\href
  {https://doi.org/10.1103/PhysRevA.65.033608} {\bibfield  {journal} {\bibinfo
  {journal} {Physical Review A}\ }\textbf {\bibinfo {volume} {65}},\ \bibinfo
  {pages} {033608} (\bibinfo {year} {2002})}\BibitemShut {NoStop}%
\bibitem [{\citenamefont {Perrin}\ \emph {et~al.}(2019)\citenamefont {Perrin},
  \citenamefont {Bidel}, \citenamefont {Zahzam}, \citenamefont {Blanchard},
  \citenamefont {Bresson},\ and\ \citenamefont
  {Cadoret}}]{perrin_proof--principle_2019}%
  \BibitemOpen
  \bibfield  {author} {\bibinfo {author} {\bibfnamefont {I.}~\bibnamefont
  {Perrin}}, \bibinfo {author} {\bibfnamefont {Y.}~\bibnamefont {Bidel}},
  \bibinfo {author} {\bibfnamefont {N.}~\bibnamefont {Zahzam}}, \bibinfo
  {author} {\bibfnamefont {C.}~\bibnamefont {Blanchard}}, \bibinfo {author}
  {\bibfnamefont {A.}~\bibnamefont {Bresson}},\ and\ \bibinfo {author}
  {\bibfnamefont {M.}~\bibnamefont {Cadoret}},\ }\href
  {https://doi.org/10.1103/PhysRevA.99.013601} {\bibfield  {journal} {\bibinfo
  {journal} {Physical Review A}\ }\textbf {\bibinfo {volume} {99}},\ \bibinfo
  {pages} {013601} (\bibinfo {year} {2019})}\BibitemShut {NoStop}%
\bibitem [{\citenamefont {Dickerson}\ \emph {et~al.}(2013)\citenamefont
  {Dickerson}, \citenamefont {Hogan}, \citenamefont {Sugarbaker}, \citenamefont
  {Johnson},\ and\ \citenamefont {Kasevich}}]{dickerson_multiaxis_2013}%
  \BibitemOpen
  \bibfield  {author} {\bibinfo {author} {\bibfnamefont {S.~M.}\ \bibnamefont
  {Dickerson}}, \bibinfo {author} {\bibfnamefont {J.~M.}\ \bibnamefont
  {Hogan}}, \bibinfo {author} {\bibfnamefont {A.}~\bibnamefont {Sugarbaker}},
  \bibinfo {author} {\bibfnamefont {D.~M.~S.}\ \bibnamefont {Johnson}},\ and\
  \bibinfo {author} {\bibfnamefont {M.~A.}\ \bibnamefont {Kasevich}},\ }\href
  {https://doi.org/10.1103/PhysRevLett.111.083001} {\bibfield  {journal}
  {\bibinfo  {journal} {Physical Review Letters}\ }\textbf {\bibinfo {volume}
  {111}},\ \bibinfo {pages} {083001} (\bibinfo {year} {2013})}\BibitemShut
  {NoStop}%
\bibitem [{\citenamefont {Sugarbaker}\ \emph {et~al.}(2013)\citenamefont
  {Sugarbaker}, \citenamefont {Dickerson}, \citenamefont {Hogan}, \citenamefont
  {Johnson},\ and\ \citenamefont {Kasevich}}]{sugarbaker_enhanced_2013}%
  \BibitemOpen
  \bibfield  {author} {\bibinfo {author} {\bibfnamefont {A.}~\bibnamefont
  {Sugarbaker}}, \bibinfo {author} {\bibfnamefont {S.~M.}\ \bibnamefont
  {Dickerson}}, \bibinfo {author} {\bibfnamefont {J.~M.}\ \bibnamefont
  {Hogan}}, \bibinfo {author} {\bibfnamefont {D.~M.~S.}\ \bibnamefont
  {Johnson}},\ and\ \bibinfo {author} {\bibfnamefont {M.~A.}\ \bibnamefont
  {Kasevich}},\ }\href {https://doi.org/10.1103/PhysRevLett.111.113002}
  {\bibfield  {journal} {\bibinfo  {journal} {Physical Review Letters}\
  }\textbf {\bibinfo {volume} {111}},\ \bibinfo {pages} {113002} (\bibinfo
  {year} {2013})}\BibitemShut {NoStop}%
\bibitem [{\citenamefont {Freier}(2017)}]{freier_atom_2017}%
  \BibitemOpen
  \bibfield  {author} {\bibinfo {author} {\bibfnamefont {C.}~\bibnamefont
  {Freier}},\ }\emph {\bibinfo {title} {Atom interferometry at geodetic
  observatories}},\ \href {https://edoc.hu-berlin.de/handle/18452/18447} {Ph.D.
  thesis},\ \bibinfo  {school} {Humboldt-Universit{\"a}t zu Berlin}, \bibinfo
  {address} {Berlin, Germany} (\bibinfo {year} {2017})\BibitemShut {NoStop}%
\bibitem [{\citenamefont {Templier}\ \emph {et~al.}(2021)\citenamefont
  {Templier}, \citenamefont {Hauden}, \citenamefont {Cheiney}, \citenamefont
  {Napolitano}, \citenamefont {Porte}, \citenamefont {Bouyer}, \citenamefont
  {Barrett},\ and\ \citenamefont
  {Battelier}}]{templier_carrier-suppressed_2021}%
  \BibitemOpen
  \bibfield  {author} {\bibinfo {author} {\bibfnamefont {S.}~\bibnamefont
  {Templier}}, \bibinfo {author} {\bibfnamefont {J.}~\bibnamefont {Hauden}},
  \bibinfo {author} {\bibfnamefont {P.}~\bibnamefont {Cheiney}}, \bibinfo
  {author} {\bibfnamefont {F.}~\bibnamefont {Napolitano}}, \bibinfo {author}
  {\bibfnamefont {H.}~\bibnamefont {Porte}}, \bibinfo {author} {\bibfnamefont
  {P.}~\bibnamefont {Bouyer}}, \bibinfo {author} {\bibfnamefont
  {B.}~\bibnamefont {Barrett}},\ and\ \bibinfo {author} {\bibfnamefont
  {B.}~\bibnamefont {Battelier}},\ }\href
  {https://doi.org/10.1103/PhysRevApplied.16.044018} {\bibfield  {journal}
  {\bibinfo  {journal} {Physical Review Applied}\ }\textbf {\bibinfo {volume}
  {16}},\ \bibinfo {pages} {044018} (\bibinfo {year} {2021})}\BibitemShut
  {NoStop}%
\bibitem [{\citenamefont {Foster}\ \emph {et~al.}(2002)\citenamefont {Foster},
  \citenamefont {Fixler}, \citenamefont {McGuirk},\ and\ \citenamefont
  {Kasevich}}]{foster_method_2002}%
  \BibitemOpen
  \bibfield  {author} {\bibinfo {author} {\bibfnamefont {G.~T.}\ \bibnamefont
  {Foster}}, \bibinfo {author} {\bibfnamefont {J.~B.}\ \bibnamefont {Fixler}},
  \bibinfo {author} {\bibfnamefont {J.~M.}\ \bibnamefont {McGuirk}},\ and\
  \bibinfo {author} {\bibfnamefont {M.~A.}\ \bibnamefont {Kasevich}},\ }\href
  {https://doi.org/10.1364/OL.27.000951} {\bibfield  {journal} {\bibinfo
  {journal} {Optics Letters}\ }\textbf {\bibinfo {volume} {27}},\ \bibinfo
  {pages} {951} (\bibinfo {year} {2002})}\BibitemShut {NoStop}%
\bibitem [{\citenamefont {DiFrancesco}\ \emph {et~al.}(2009)\citenamefont
  {DiFrancesco}, \citenamefont {Grierson}, \citenamefont {Kaputa},\ and\
  \citenamefont {Meyer}}]{difrancesco_gravity_2009}%
  \BibitemOpen
  \bibfield  {author} {\bibinfo {author} {\bibfnamefont {D.}~\bibnamefont
  {DiFrancesco}}, \bibinfo {author} {\bibfnamefont {A.}~\bibnamefont
  {Grierson}}, \bibinfo {author} {\bibfnamefont {D.}~\bibnamefont {Kaputa}},\
  and\ \bibinfo {author} {\bibfnamefont {T.}~\bibnamefont {Meyer}},\ }\href
  {https://doi.org/10.1111/j.1365-2478.2008.00764.x} {\bibfield  {journal}
  {\bibinfo  {journal} {Geophysical Prospecting}\ }\textbf {\bibinfo {volume}
  {57}},\ \bibinfo {pages} {615} (\bibinfo {year} {2009})}\BibitemShut
  {NoStop}%
\bibitem [{\citenamefont {Giltner}\ \emph {et~al.}(1995)\citenamefont
  {Giltner}, \citenamefont {McGowan},\ and\ \citenamefont
  {Lee}}]{giltner_atom_1995}%
  \BibitemOpen
  \bibfield  {author} {\bibinfo {author} {\bibfnamefont {D.~M.}\ \bibnamefont
  {Giltner}}, \bibinfo {author} {\bibfnamefont {R.~W.}\ \bibnamefont
  {McGowan}},\ and\ \bibinfo {author} {\bibfnamefont {S.~A.}\ \bibnamefont
  {Lee}},\ }\href {https://doi.org/10.1103/PhysRevLett.75.2638} {\bibfield
  {journal} {\bibinfo  {journal} {Physical Review Letters}\ }\textbf {\bibinfo
  {volume} {75}},\ \bibinfo {pages} {2638} (\bibinfo {year}
  {1995})}\BibitemShut {NoStop}%
\bibitem [{\citenamefont {Chiow}\ \emph {et~al.}(2011)\citenamefont {Chiow},
  \citenamefont {Kovachy}, \citenamefont {Chien},\ and\ \citenamefont
  {Kasevich}}]{chiow_102ensuremathhbark_2011}%
  \BibitemOpen
  \bibfield  {author} {\bibinfo {author} {\bibfnamefont {S.-w.}\ \bibnamefont
  {Chiow}}, \bibinfo {author} {\bibfnamefont {T.}~\bibnamefont {Kovachy}},
  \bibinfo {author} {\bibfnamefont {H.-C.}\ \bibnamefont {Chien}},\ and\
  \bibinfo {author} {\bibfnamefont {M.~A.}\ \bibnamefont {Kasevich}},\ }\href
  {https://doi.org/10.1103/PhysRevLett.107.130403} {\bibfield  {journal}
  {\bibinfo  {journal} {Physical Review Letters}\ }\textbf {\bibinfo {volume}
  {107}},\ \bibinfo {pages} {130403} (\bibinfo {year} {2011})}\BibitemShut
  {NoStop}%
\bibitem [{\citenamefont {McDonald}\ \emph {et~al.}(2013)\citenamefont
  {McDonald}, \citenamefont {Kuhn}, \citenamefont {Bennetts}, \citenamefont
  {Debs}, \citenamefont {Hardman}, \citenamefont {Johnsson}, \citenamefont
  {Close},\ and\ \citenamefont {Robins}}]{mcdonald_80ensuremathhbark_2013}%
  \BibitemOpen
  \bibfield  {author} {\bibinfo {author} {\bibfnamefont {G.~D.}\ \bibnamefont
  {McDonald}}, \bibinfo {author} {\bibfnamefont {C.~C.~N.}\ \bibnamefont
  {Kuhn}}, \bibinfo {author} {\bibfnamefont {S.}~\bibnamefont {Bennetts}},
  \bibinfo {author} {\bibfnamefont {J.~E.}\ \bibnamefont {Debs}}, \bibinfo
  {author} {\bibfnamefont {K.~S.}\ \bibnamefont {Hardman}}, \bibinfo {author}
  {\bibfnamefont {M.}~\bibnamefont {Johnsson}}, \bibinfo {author}
  {\bibfnamefont {J.~D.}\ \bibnamefont {Close}},\ and\ \bibinfo {author}
  {\bibfnamefont {N.~P.}\ \bibnamefont {Robins}},\ }\href
  {https://doi.org/10.1103/PhysRevA.88.053620} {\bibfield  {journal} {\bibinfo
  {journal} {Physical Review A}\ }\textbf {\bibinfo {volume} {88}},\ \bibinfo
  {pages} {053620} (\bibinfo {year} {2013})}\BibitemShut {NoStop}%
\bibitem [{\citenamefont {Rodzinka}\ \emph {et~al.}(2024)\citenamefont
  {Rodzinka}, \citenamefont {Dionis}, \citenamefont {Calmels}, \citenamefont
  {Beldjoudi}, \citenamefont {Béguin}, \citenamefont {Guéry-Odelin},
  \citenamefont {Allard}, \citenamefont {Sugny},\ and\ \citenamefont
  {Gauguet}}]{rodzinka_optimal_2024}%
  \BibitemOpen
  \bibfield  {author} {\bibinfo {author} {\bibfnamefont {T.}~\bibnamefont
  {Rodzinka}}, \bibinfo {author} {\bibfnamefont {E.}~\bibnamefont {Dionis}},
  \bibinfo {author} {\bibfnamefont {L.}~\bibnamefont {Calmels}}, \bibinfo
  {author} {\bibfnamefont {S.}~\bibnamefont {Beldjoudi}}, \bibinfo {author}
  {\bibfnamefont {A.}~\bibnamefont {Béguin}}, \bibinfo {author} {\bibfnamefont
  {D.}~\bibnamefont {Guéry-Odelin}}, \bibinfo {author} {\bibfnamefont
  {B.}~\bibnamefont {Allard}}, \bibinfo {author} {\bibfnamefont
  {D.}~\bibnamefont {Sugny}},\ and\ \bibinfo {author} {\bibfnamefont
  {A.}~\bibnamefont {Gauguet}},\ }\href
  {https://doi.org/10.1038/s41467-024-54539-w} {\bibfield  {journal} {\bibinfo
  {journal} {Nature Communications}\ }\textbf {\bibinfo {volume} {15}},\
  \bibinfo {pages} {10281} (\bibinfo {year} {2024})}\BibitemShut {NoStop}%
\bibitem [{\citenamefont {Szigeti}\ \emph {et~al.}(2012)\citenamefont
  {Szigeti}, \citenamefont {Debs}, \citenamefont {Hope}, \citenamefont
  {Robins},\ and\ \citenamefont {Close}}]{szigeti_why_2012}%
  \BibitemOpen
  \bibfield  {author} {\bibinfo {author} {\bibfnamefont {S.~S.}\ \bibnamefont
  {Szigeti}}, \bibinfo {author} {\bibfnamefont {J.~E.}\ \bibnamefont {Debs}},
  \bibinfo {author} {\bibfnamefont {J.~J.}\ \bibnamefont {Hope}}, \bibinfo
  {author} {\bibfnamefont {N.~P.}\ \bibnamefont {Robins}},\ and\ \bibinfo
  {author} {\bibfnamefont {J.~D.}\ \bibnamefont {Close}},\ }\href
  {https://doi.org/10.1088/1367-2630/14/2/023009} {\bibfield  {journal}
  {\bibinfo  {journal} {New Journal of Physics}\ }\textbf {\bibinfo {volume}
  {14}},\ \bibinfo {pages} {023009} (\bibinfo {year} {2012})}\BibitemShut
  {NoStop}%
\bibitem [{\citenamefont {Saywell}\ \emph {et~al.}(2023)\citenamefont
  {Saywell}, \citenamefont {Carey}, \citenamefont {Light}, \citenamefont
  {Szigeti}, \citenamefont {Milne}, \citenamefont {Gill}, \citenamefont {Goh},
  \citenamefont {Perunicic}, \citenamefont {Wilson}, \citenamefont {Macrae},
  \citenamefont {Rischka}, \citenamefont {Everitt}, \citenamefont {Robins},
  \citenamefont {Anderson}, \citenamefont {Hush},\ and\ \citenamefont
  {Biercuk}}]{saywell_enhancing_2023}%
  \BibitemOpen
  \bibfield  {author} {\bibinfo {author} {\bibfnamefont {J.~C.}\ \bibnamefont
  {Saywell}}, \bibinfo {author} {\bibfnamefont {M.~S.}\ \bibnamefont {Carey}},
  \bibinfo {author} {\bibfnamefont {P.~S.}\ \bibnamefont {Light}}, \bibinfo
  {author} {\bibfnamefont {S.~S.}\ \bibnamefont {Szigeti}}, \bibinfo {author}
  {\bibfnamefont {A.~R.}\ \bibnamefont {Milne}}, \bibinfo {author}
  {\bibfnamefont {K.~S.}\ \bibnamefont {Gill}}, \bibinfo {author}
  {\bibfnamefont {M.~L.}\ \bibnamefont {Goh}}, \bibinfo {author} {\bibfnamefont
  {V.~S.}\ \bibnamefont {Perunicic}}, \bibinfo {author} {\bibfnamefont {N.~M.}\
  \bibnamefont {Wilson}}, \bibinfo {author} {\bibfnamefont {C.~D.}\
  \bibnamefont {Macrae}}, \bibinfo {author} {\bibfnamefont {A.}~\bibnamefont
  {Rischka}}, \bibinfo {author} {\bibfnamefont {P.~J.}\ \bibnamefont
  {Everitt}}, \bibinfo {author} {\bibfnamefont {N.~P.}\ \bibnamefont {Robins}},
  \bibinfo {author} {\bibfnamefont {R.~P.}\ \bibnamefont {Anderson}}, \bibinfo
  {author} {\bibfnamefont {M.~R.}\ \bibnamefont {Hush}},\ and\ \bibinfo
  {author} {\bibfnamefont {M.~J.}\ \bibnamefont {Biercuk}},\ }\href
  {https://doi.org/10.1038/s41467-023-43374-0} {\bibfield  {journal} {\bibinfo
  {journal} {Nature Communications}\ }\textbf {\bibinfo {volume} {14}},\
  \bibinfo {pages} {7626} (\bibinfo {year} {2023})}\BibitemShut {NoStop}%
\end{thebibliography}

\end{document}